\documentclass[%twocolumn,%showpacs,preprintnumbers,
amsmath,amssymb,%aps,
prd,nofootinbib,floatfix,12pt,%preprint,11pt
]{revtex4}

\usepackage{amsmath,amssymb,amsfonts}
\usepackage[dvips]{color}
\usepackage{xcolor}
\usepackage{dsfont}
\usepackage{mathrsfs}
\usepackage{hyperref}
\usepackage{float}
\usepackage{graphicx}
\usepackage{graphics}
\usepackage{setspace}
\usepackage{lmodern}

\hypersetup{colorlinks,citecolor=blue,urlcolor=blue, linkcolor=black}
\newcommand\beq{\begin{eqnarray}}
\newcommand\eeq{\end{eqnarray}}

\newcommand{\nocontentsline}[3]{}
\newcommand{\tocless}[2]{\bgroup\let\addcontentsline=\nocontentsline#1{#2}\egroup}

\begin{document}
\renewcommand{\theequation}{\arabic{section}.\arabic{equation}}
\renewcommand{\thefigure}{\arabic{section}.\arabic{figure}}
\renewcommand{\thetable}{\arabic{section}.\arabic{table}}

\title{\large \baselineskip=20pt 
%Implications for the extended Higgs sectors with universal suppression and invisible decays for the 125 GeV Higgs}
The depleted Higgs boson: searches for universal coupling suppression, invisible decays, and mixed-in scalars}

\author{Prudhvi N.~Bhattiprolu and James D.~Wells}
\affiliation{\it Leinweber Center for Theoretical Physics,\\
University of Michigan, Ann Arbor MI 48109, USA}

\begin{abstract}\normalsize \baselineskip=15.5pt
%There are two simple ways that the standard signals of the Standard Model Higgs boson can be depleted
Two simple ways by which the standard signals of the Standard Model Higgs boson can be depleted are:
its couplings to fermions and gauge bosons can be suppressed by a universal factor, and part of its branching fraction can be drained into invisible final states. A large class of theories can impose one or both of these depletion factors, even if mild, by way of  additional scalar bosons that are singlets under the Standard Model but mix with the Higgs boson. We perform a comprehensive survey of the present status of the depleted Higgs boson, and discuss future prospects for detecting the presence of either depletion factor. We also survey the constraints status and future detection prospects for the generic case of extra mixed-in scalars which generically lead to these depletion factors for the Higgs boson. We find, for example, that precision study of the Higgs boson in many cases is more powerful than searches for the extra scalar states, given the slate of next-generation experiments that are
on the horizon.
\end{abstract}

\maketitle
\vspace{-1.1cm}

% disable subsections and subsubsections in the TOC
\makeatletter
\def\l@subsubsection#1#2{}
\makeatother

\baselineskip=18pt

\tableofcontents

\baselineskip=14.55pt

\setcounter{footnote}{1}
\setcounter{figure}{0}
\setcounter{table}{0}

\newpage
\section{Introduction\label{sec:introduction}}
\setcounter{equation}{0}
\setcounter{figure}{0}
\setcounter{table}{0}
\setcounter{footnote}{1}

One of the biggest triumphs of the Standard Model (SM) is the explanation of
the electroweak symmetry breaking via a CP-even complex scalar $H \sim ({\bf 1}, {\bf 2}, \frac{1}{2})$
which further predicts the existence of a neutral Higgs boson $h_{\rm SM}$ that is part of the neutral component
of the $SU(2)_L$ doublet $H$.
So far all the measured properties of the $\sim 125$ GeV physical Higgs boson that was discovered in 2012
at the Large Hadron Collider (LHC) are consistent with the SM predictions and therefore
the observed Higgs boson $h$ is often identified as the SM Higgs boson $h_{\rm SM}$.

It is important to note, however, that as the Higgs measurements get more precise there is a
well-motivated possibility that the observations can deviate from the SM predictions thereby pointing towards
the presence of a more complicated Higgs sector.
In this paper we study the prospect of the observed Higgs boson $h$ having a universal depletion factor $\delta$,
suppressing all its couplings to the SM states, along with an invisible width.\footnote{See also refs.\,\cite{Biekotter:2022ckj,Fernandez-Martinez:2022stj} for related studies on the impact of two universal depletion factors on Higgs signal strength measurements.} The invisible width constitutes a second depletion factor for the Higgs boson because its standard decay final state branching fractions are depleted due to the additional invisible mode.

To set notation, the SM Higgs boson $h_{\rm SM}$ is related to the observed Higgs boson $h$ in the
following way:
\beq
h_{\rm SM} &=& \sqrt{1 - \delta^2} \, h + \cdots,
\label{eq:hSM_h_relation}
\eeq
where $\cdots$ include contributions from additional physical Higgs states that can arise from extended Higgs
sectors. In other words, we identify the label ``Higgs boson" $h$ with the $h125$ discovered resonance and not necessarily with the Platonic ideal Standard Model Higgs boson $h_{\rm SM}$. 
Below, we obtain constraints on the
universal depletion factor $\delta$, dependent on the invisible width of $h$, from various searches
for the $\sim$125 GeV Higgs boson at the LHC while
remaining agnostic to the particulars of the extended Higgs sectors.
We also later reinterpret some projections at the International Linear Collider (ILC) and
the High Luminosity LHC (HL-LHC) from the invisible decays of $h$.
We emphasize that the relation in eq.~(\ref{eq:hSM_h_relation}) can naturally arise from many
extended Higgs sectors.

As a concrete example for an extended Higgs sector that leads to a universal depletion factor $\delta$
and an invisible width for $h$, we first consider an extension to the SM with a real scalar
$S \sim ({\bf 1}, {\bf 1}, 0)$ that mixes with the SM Higgs boson $h_{\rm SM}$ to give rise to two physical
mass eigenstates: the observed Higgs $h$ and an exotic Higgs $\phi$.
We take the singlet scalar $S$ to only have invisible decays which translates to an invisible width for $h$
via its mixing with $S$.
SM extensions with a singlet scalar were extensively studied in the literature,
see e.g.\ refs.\,\cite{McDonald:1993ex,Martin:1999qf,Wells:2002gq,Schabinger:2005ei,Patt:2006fw,OConnell:2006rsp,Bowen:2007ia,Profumo:2007wc,Dawson:2009yx,Espinosa:2011ax,Pruna:2013bma,Lopez-Val:2014jva,Chen:2014ask,Robens:2015gla,Dawson:2015haa,Robens:2016xkb,Ilnicka:2018def,Dawson:2021jcl,Falkowski:2015iwa,Barger:2007im,Bertolini:2012gu} and the references therein. 

For illustration purposes, we only restrict ourselves to the mass of the exotic Higgs $\phi$  to be in between
$m_h/2$ and $2 m_h$ such that the decays $h \rightarrow \phi \phi$ and $\phi \rightarrow h h$ are
kinematically forbidden.
As we will see below that after re-examining the parameter space with the latest bounds,
the indirect constraints from the precision probes for the observed Higgs boson $h$ at the LHC
are the dominant ones in general compared to the bounds from the searches for additional
neutral Higgs states, and from the oblique parameters $S, T,$ and $U$.
This is in accord with the findings in refs.\,\cite{Robens:2015gla,Robens:2016xkb} for
$m_h/2 \le m_\phi \le 2 m_h$ specifically for the case where $h$ does not have any invisible decays.

We also show some future projections at the ILC.
As a further extension to this case, we also briefly consider the case with $N$ such scalars that are
assumed to mix equally among themselves, and with the SM Higgs boson. Once again, as we will see below,
we find the indirect bounds from precision probes for the observed Higgs boson $h$
moderately constrain the parameter space,
which gets stronger as $N$ and/or the invisible width of $h$ gets larger.

The remaining sections of this paper are structured as follows.
In Section~\ref{sec:125GeVHiggs}, we define ``the depleted Higgs boson" -- Higgs boson with a universal suppression and invisible width -- and obtain its production cross-sections, total width, and branching ratios relative to that of the 125 GeV Higgs in the SM. 
In Section~\ref{sec:SM_Higgs_bounds} we study the constraints on the universal depletion factor $\delta$,
that varies with the invisible width of $h$, from the latest LHC precision probes for the $\sim 125$ GeV Higgs boson
in various decay channels. We also consider some example future projections from ILC and HL-LHC.
In Section~\ref{sec:scalar_ext}, we consider the SM extension with a real singlet scalar,
and in Section~\ref{sec:Singlet_Higgs_bounds} we reexamine the present direct and indirect bounds along with
some future projections from collider searches, and the current constraints from
the Peskin-Takeuchi $S, T,$ and $U$ parameters.
We also briefly consider a Higgs sector with $N$ real singlet scalars, under some simplifying assumptions,
and study the implications of direct and indirect searches on such a scenario in Section~\ref{sec:NSinglets_Higgs_bounds}.
Finally, we end with some concluding remarks in Section~\ref{sec:conclusion}.

%%%%%%%%%%%%%%%%%%%%%%%%%%%%%%%%%%%%%%%
%%%%%%%%%%%%%%%%%%%%%%%%%%%%%%%%%%%%%%%
%\section{125 GeV Higgs with a universal depletion factor and an invisible width\label{sec:125GeVHiggs}}
\section{Higgs boson with a universal suppression and invisible width\label{sec:125GeVHiggs}}
\setcounter{equation}{0}
\setcounter{figure}{0}
\setcounter{table}{0}
\setcounter{footnote}{1}

Consider the Higgs boson $h$ with its mass of $\sim$125 GeV
%that is discovered in 2012 by the ATLAS and CMS collaborations at the Large Hadron Collider (LHC)
and with a universal depletion factor $\delta$ for all its couplings to the SM states,
such that
the production cross-sections of $h$ are given by
\beq
\sigma^h &=& (1 - \delta^2) \, \sigma^{125}_{\rm SM},
\label{eq:sigma}
\eeq
where $\sigma^{125}_{\rm SM}$ are the production cross-sections for 125 GeV Higgs in the SM.
Additionally, assume that the 125 GeV Higgs also has an invisible width that is parameterized as
$\delta^2 \, \Gamma_{\rm inv}$.
Therefore, the total width of $h$ is
\beq
\Gamma^h &=& (1 - \delta^2) \, \Gamma^{125}_{\rm SM} + \delta^2 \, \Gamma_{\rm inv},
\label{eq:Gamma_125_def}
\eeq
where $\Gamma^{125}_{\rm SM}$ stands for the total width of the 125 GeV
Higgs boson in the SM.
Upon defining
\beq
\label{eq:kappainv}
\kappa_{\rm inv} &\equiv& \Gamma_{\rm inv}/\Gamma^{125}_{\rm SM},
\eeq
eq.~(\ref{eq:Gamma_125_def}) can be reexpressed as
\beq
\frac{\Gamma^h}{\Gamma^{125}_{\rm SM}} &=& 1 - (1 - \kappa_{\rm inv}) \delta^2.
\label{eq:SM_Gammah}
\eeq

Given these definitions, the ratio of the branching ratio $B^h_j$ of $h \rightarrow j^{\rm th}$ SM final state to that of the
corresponding branching ratio in the SM $B^{125}_{{\rm SM}, j}$, and the invisible branching ratio $B^h_{\rm inv}$
can be entirely expressed in terms of just two parameters $-$ $\delta$ and $\kappa_{\rm inv}$:
\beq
\frac{B^h_j}{B^{125}_{{\rm SM}, j}} &=& \frac{1 - \delta^2}{1 - (1 - \kappa_{\rm inv}) \delta^2},
\label{eq:BSMj}\\
B^h_{\rm inv} &=& \frac{\delta^2 \kappa_{\rm inv}}{1 - (1 - \kappa_{\rm inv}) \delta^2}.
\label{eq:Binv}
\eeq
Note that in the limit $\delta \rightarrow 0$ the SM is completely recovered
(independent of $\kappa_{\rm inv}$ or $\Gamma_{\rm inv}$).

Although, at this point, the appearance of the terms $\delta$ and $\Gamma_{\rm inv}$
(or equivalently $\kappa_{\rm inv}$) might seem ad-hoc, we later consider the case
where the SM is extended by a real scalar $S \sim ({\bf 1}, {\bf 1}, 0)$ that decays
invisibly and mixes with the SM Higgs boson $h_{\rm SM}$. In which case, $\delta$
will be the sine of the mixing angle between $h_{\rm SM}$ and $S$, and
$\Gamma_{\rm inv}$ is the invisible decay width of $S$.
More generally, the terms $\delta$ and $\kappa_{\rm inv}$ naturally appear
in many models beyond the SM, since these ``singlets" can represent any number of possible states with exotic charges in a sector beyond the SM.
Furthermore, we will see that the constraints on $\delta$ and $\kappa_{\rm inv}$
that come just from the precision probes for the 125 GeV SM Higgs boson moderately constrain
the extensions to the Higgs sector.
The case with $N \ge 1$ real singlet Higgs boson(s) that we later consider will illustrate that.

%%%%%%%%%%%%%%%%%%%%%%%%%%%%%%%%%%%%%%%
%%%%%%%%%%%%%%%%%%%%%%%%%%%%%%%%%%%%%%%
%%%%%%%%%%%%%%%%%%%%%%%%%%%%%%%%%%%%%%%

%\subsection{Constraints on ${(\delta, \kappa_{\rm inv})}$ \label{sec:SM_Higgs_bounds}}
\section{Constraints on depletion factors $\delta$ and $\kappa_{\rm inv}$\label{sec:SM_Higgs_bounds}}
\setcounter{equation}{0}
\setcounter{figure}{0}
\setcounter{table}{0}
\setcounter{footnote}{1}

We now turn to constraints on the 2-dimensional parameter space of
$(\delta, \kappa_{\rm inv})$ that come from searches by ATLAS and CMS collaborations at the
LHC for the observed Higgs boson decaying invisibly or to the SM final states.
From invisible searches, a bound is usually reported on the invisible branching ratio $B_{\rm inv}$ of
the Higgs boson multiplied with its production cross-section 
relative to the production cross-section in the SM $\sigma/\sigma_{\rm SM}$.
In particular, an upper bound is reported on $\mu_{\rm inv}$\footnote[2]{For convenience, we define the invisible branching ratio of
the Higgs boson multiplied with its production cross-section 
relative to the production cross-section in the SM as $\mu_{\rm inv}$, which is not to be confused with
the signal strength modifier that is defined in eq.~(\ref{eq:signalstrength}).}
which we define as
\beq
\mu_{\rm inv} &\equiv& \frac{\sigma}{\sigma_{\rm SM}} B_{\rm inv},
\label{eq:mu_inv}
\eeq
which in our case can be written as (see eq.~(\ref{eq:sigma})) 
\beq
\mu_{\rm inv} &=& (1 - \delta^2) B^h_{\rm inv}.
\eeq
Therefore, using eq.~(\ref{eq:Binv}), the above equation imposes the following constraint
on $(\delta, \kappa_{\rm inv})$ parameter space
\beq
\frac{\delta^2 \, (1 - \delta^2) \, \kappa_{\rm inv}}{1 - (1 - \kappa_{\rm inv}) \delta^2} &<& \mu_{\rm inv}.
\eeq
Table~\ref{tab:Higgs125_inv} shows some recent LHC upper bounds on the quantity $\mu_{\rm inv}$ at 95\% CL by ATLAS and CMS experiments.
%%%%%%%%%%%%%%%%%%%%%%%%%%%%%%%%%%%%%%%%%%%%%%%%%%%%%%%%%%%%%%%%%%%%%%%%%%%%%%%%%%%%%%%%%%%
\begin{table}
\begin{minipage}[]{0.95\linewidth}
\caption{Recent LHC searches for invisible decays of the 125 GeV Higgs boson by ATLAS and CMS experiments along with the reported 95\% CL upper bounds on $\mu_{\rm inv}$.
$\mu_{\rm inv}$ is defined to be the product of the invisible branching fraction of the Higgs boson and its production cross-section relative to that of the SM.
For convenience, various searches are assigned labels which will be used in the figure(s) below.
\label{tab:Higgs125_inv}}
\end{minipage}
\begin{center}
\begin{tabular}{|c | c | c |}
\hline
~Label~
%& ~$\left(\sqrt{s} \text{ in TeV}, \int \mathcal{L} \, dt \text{ in {fb$^{-1}$}}\right)$~
&  ~95\% CL upper bound on $\mu_{\rm inv}$~  & ~Reference~\\
\hline
\hline
~ATLAS 2020 ($h \rightarrow \text{inv}$)~ 
%&
%~(13, 139), (8, 20.3), (7, 4.7)~ 
&  
~0.11~
&
~\cite{ATLAS:2020kdi}~
\\[1pt]
\hline
~ATLAS 2020 ($h \rightarrow \text{inv}$, 13 TeV)~ 
%&
%~(13, 139)~ 
&  
~0.13~
&
~\cite{ATLAS:2020kdi}~
\\[1pt]
\hline
~CMS 2018 ($h \rightarrow \text{inv}$)~
%&
%~(13, 38.2), (8, 19.7), (7, 4.9)~ 
&  
~0.18~
&
~\cite{CMS:2018yfx}~
\\[1pt]
\hline
~CMS 2018 ($h \rightarrow \text{inv}$, 13 TeV)~ 
%&
%~(13, 38.2)~ 
&  
~0.33~
&
~\cite{CMS:2018yfx}~
\\[1pt]
\hline
\end{tabular}
\end{center}
\end{table}
%%%%%%%%%%%%%%%%%%%%%%%%%%%%%%%%%%%%%%%%%%%%%%%%%%%%%%%%%%%%%%%%%%%%%%%%%%%%%%%%%%%%%%%%%%%
%%%%%%%%%%%%%%%%%%%%%%%%%%%%%%%%%%%%%%%%%%%%%%%%%%%%%%%%%%%%%%%%%%%%%%%%%%%%%%%%%%%%%%%%%%%
\begin{table}
\begin{minipage}[]{0.95\linewidth}
\caption{Recent LHC precision probes for the 125 GeV Higgs boson decaying into the (visible) SM final states by ATLAS and CMS experiments along with the measured signal strength modifier $\mu$ and its overall uncertainty as reported by ATLAS/CMS that accounts for various systematic and statistical uncertainties.
For convenience, various searches are assigned labels which will be used in the figure(s) below.
\label{tab:Higgs125_SM}}
\end{minipage}
\begin{center}
\begin{tabular}{|c | c | c | c |}
\hline
~Label~
% & ~$\left(\sqrt{s} \text{ in TeV}, \int \mathcal{L} \, dt \text{ in {fb$^{-1}$}}\right)$~
&  ~Observed $\mu$~  & ~Reference~\\
\hline
\hline
~CMS 2021 ($h \rightarrow \gamma \gamma$)~ 
&  
~$1.12 \pm 0.09$~
&
~\cite{CMS:2021kom}~
\\[1pt]
\hline
~ATLAS 2022 ($h \rightarrow \gamma \gamma$)~ 
&  
~$1.04^{+0.10}_{-0.09}$~
&
~\cite{ATLAS:2022tnm}~
\\[1pt]
\hline
~ATLAS 2020 ($h \rightarrow Z Z \rightarrow 4\ell$)~ 
&  
~$1.01 \pm 0.11$~
&
~\cite{ATLAS:2020rej}~
\\[1pt]
\hline
~CMS 2021 ($h \rightarrow Z Z \rightarrow 4\ell$)~ 
&  
~$0.94 \pm 0.11$~
&
~\cite{CMS:2021ugl}~
\\[1pt]
\hline
~ATLAS 2022 ($h \rightarrow W W \rightarrow e \nu \mu \nu$)~ 
&  
~$1.09 \pm 0.11$~
&
~\cite{ATLAS:2022ooq}~
\\[1pt]
\hline
~CMS 2020 ($h \rightarrow W W \rightarrow e \nu \mu \nu$)~ 
&  
~$1.05 \pm 0.12$~
&
~\cite{CMS:2020dvg}~
\\[1pt]
\hline
~CMS 2018 ($h \rightarrow b b$)~ 
&  
~$1.04 \pm 0.20$~
&
~\cite{CMS:2018nsn}~
\\[1pt]
\hline
~ATLAS 2018 ($h \rightarrow b b$)~ 
&  
~$1.01 \pm 0.20$~
&
~\cite{ATLAS:2018kot}~
\\[1pt]
\hline
~ATLAS 2022 ($h \rightarrow \tau \tau$)~ 
&  
~$0.93^{+0.13}_{-0.12}$~
&
~\cite{ATLAS:2022yrq}~
\\[1pt]
\hline
~CMS 2022 ($h \rightarrow \tau \tau$)~ 
&  
~$0.82 \pm 0.11$~
&
~\cite{CMS:2022kdi}~
\\[1pt]
\hline
~CMS 2020 ($h \rightarrow \mu \mu$)~ 
&  
~$1.19^{+0.44}_{-0.42}$~
&
~\cite{CMS:2020xwi}~
\\[1pt]
\hline
~ATLAS 2020 ($h \rightarrow \mu \mu$)~ 
&  
~$1.2 \pm 0.6$~
&
~\cite{ATLAS:2020fzp}~
\\[1pt]
\hline
\end{tabular}
\end{center}
\end{table}
%%%%%%%%%%%%%%%%%%%%%%%%%%%%%%%%%%%%%%%%%%%%%%%%%%%%%%%%%%%%%%%%%%%%%%%%%%%%%%%%%%%%%%%%%%%
%%%%%%%%%%%%%%%%%%%%%%%%%%%%%%%%%%%%%%%%%%%%%%%%%%%%%%%%%%%%%%%%%%%%%%%%%%%%%%%%%%
\begin{figure}
  \begin{center}\begin{minipage}[]{0.95\linewidth}
    \includegraphics[width=15cm]{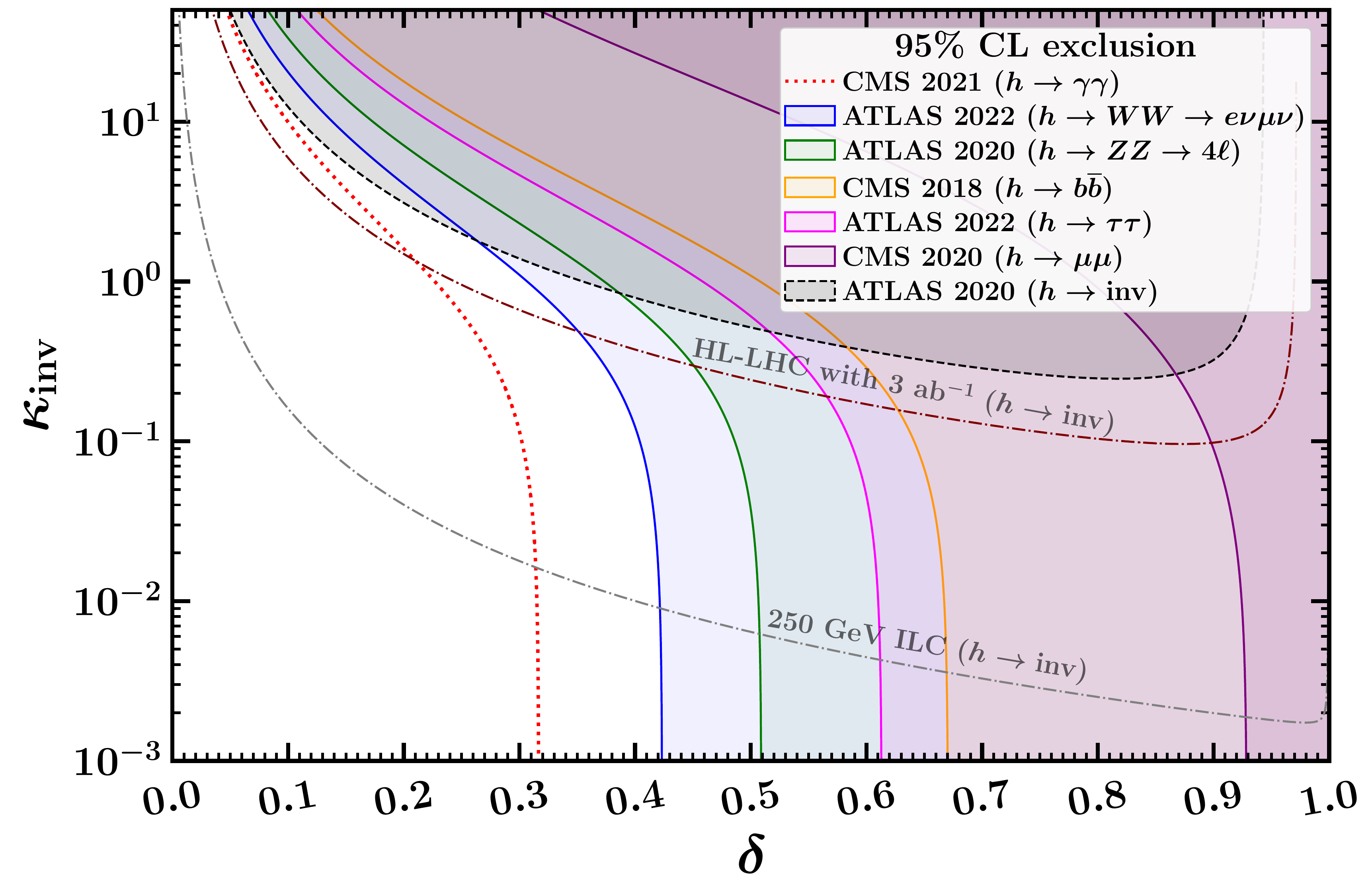}
  \end{minipage}\end{center}
 \caption{Current 95\% CL bounds on the parameter space of $(\delta, \kappa_{\rm inv})$ from
 ATLAS and CMS precision probes for 125 GeV Higgs boson in various search channels. Out of the bounds
 obtained from the reported values of $\mu_{\rm inv}$ (defined in eq.~(\ref{eq:mu_inv})) and the signal strength modifier $\mu$ from Tables~\ref{tab:Higgs125_inv} and \ref{tab:Higgs125_SM}, respectively, only the strongest bounds in each search channel are shown here.
 The shaded regions with dashed and solid borders show the bounds obtained from the searches
for $h$ to invisible and SM final states (excluding $h \rightarrow \gamma \gamma$), respectively.
The bound from $h \rightarrow \gamma \gamma$ is shown by a dotted line to emphasize that the process is loop-induced and therefore more sensitive to new physics contributions.
The dash-dotted lines show the projected 95\% CL sensitivity for invisible $h$ decays
at 250 GeV ILC and high luminosity LHC with nominal running assumptions as discussed in the text.
\label{fig:h_bounds}}
\end{figure}
%%%%%%%%%%%%%%%%%%%%%%%%%%%%%%%%%%%%%%%%%%%%%%%%%%%%%%%%%%%%%%%%%%%%%%%%%%%%%%%%%% 

As for visible final states, precision probes for the decay of the 125 GeV Higgs boson to a $j^{\rm th}$ SM final state
often report the measured signal strength modifier $\mu$ defined as
% \beq
% \mu^{+\delta_1 \ +\delta_3 \ +\ldots}_{-\delta_2 \ -\delta_4 \ -\ldots}
% &=&
% \frac{\sigma \, B_j}{\sigma_{\rm SM} \, B_{{\rm SM}, j}},
% \label{eq:signalstrength}
% \eeq
\beq
\mu^{+\sigma^+_\mu}_{-\sigma^-_\mu}
&=&
\frac{\sigma \, B_j}{\sigma_{\rm SM} \, B_{{\rm SM}, j}},
\label{eq:signalstrength}
\eeq
where $+\sigma^+_\mu$, $-\sigma^-_\mu$ account for various systematic and statistical uncertainties
on $\mu$.
In our case, the above equation also provides a constraint on
$(\delta, \kappa_{\rm inv})$ parameter space (see eqs.~(\ref{eq:sigma}) and (\ref{eq:BSMj})), which we
obtain by solving for $\delta$ as a function a $\kappa_{\rm inv}$ (or vice versa) from
% \beq
% \mu - \delta_2 - \delta_4 - \ldots
% <
% \frac{(1 - \delta^2)^2}{1 - (1 - \kappa_{\rm inv})\delta^2}
% <
% \mu + \delta_1 + \delta_3 + \ldots.
% \eeq
\beq
\chi^2
&=& \frac{1}{\sigma_\mu^2} \left (
\mu
-
\frac{(1 - \delta^2)^2}{1 - (1 - \kappa_{\rm inv})\delta^2}
\right )^2.
\eeq
Here, we require $\chi^2 = 5.99$ which corresponds to a fit with 95\% confidence level with 2 degrees
of freedom \cite{ParticleDataGroup:2020ssz}, and we simply take $\sigma_\mu = (\sigma^+_\mu + \sigma^-_\mu)/2$ (as we will be concerned here with the cases where the asymmetry, $(\sigma^+_\mu - \sigma^-_\mu)/(\sigma^+_\mu + \sigma^-_\mu)$, is either small or absent).
% Note that in order to obtain the most conservative lower (upper) bound on $\mu$, we simply subtract
% (add) the uncertainties $\delta_2, \delta_4, \ldots$ ($\delta_1, \delta_3, \ldots$) from (to) $\mu$.
Table~\ref{tab:Higgs125_SM} lists the observed $\mu$ from the recent LHC searches by ATLAS and CMS experiments for the decays of the 125 GeV Higgs boson to various SM final states.

Figure~\ref{fig:h_bounds} shows the most recent bounds from the LHC precision probes for the 125 GeV Higgs boson $h$
in various decay channels at 95\% CL. In each decay channel of $h$, only the strongest bound is shown in the figure.
The shaded regions with dashed and solid borders show the bounds obtained from the searches
for $h$ to invisible and SM final states (excluding $h \rightarrow \gamma \gamma$), respectively.
The bound obtained from the search for $h \rightarrow \gamma \gamma$ is shown by a dotted line.
Here, a dotted line is used to emphasize that the $h \rightarrow \gamma \gamma$ is a loop-induced process, and is therefore more sensitive to new physics contributions.

The future searches,
see e.g.\ 
refs.\,\cite{Fujii:2015jha,Liu:2016zki,deBlas:2019rxi,AlexanderAryshev:2022pkx,Bernardi:2022hny,deBlas:2022aow,Forslund:2022xjq,Potter:2022shg},
for $h$ further constrain this parameter space, and as candidate projections, we show the dash-dotted lines corresponding to the 95\% CL expected sensitivity for the invisible decays of $h$ at the High Luminosity LHC (HL-LHC) with $3$ ab$^{-1}$ of integrated luminosity
\cite{Liu:2016zki} and at the International Linear Collider (ILC) with Silicon detector
at $\sqrt s = 250$ GeV with an integrated luminosity of $(0.9, 0.9)$ ab$^{-1}$
for $(e^-_L e^+_R, e^-_R e^+_L)$ respectively and 
beam polarization of $(80, 30)$\% for $(e^-, e^+)$ respectively \cite{Potter:2022shg}.

Therefore, if the 125 GeV Higgs boson
has a universal suppression for all its couplings to SM fermions and gauge bosons along
with an invisible width, it is evident from Figure~\ref{fig:h_bounds} that
this scenario is moderately constrained from the current LHC searches.
And, the future precision studies of the already discovered Higgs boson should be able to
further constrain or find evidence for such a scenario with depletion factors
$(\delta, \kappa_{\rm inv})$.
Thus the precision probes for the 125 GeV Higgs boson provide for an indirect, and often the most powerful probe
(as we will see below for few examples), for a large class of theories with depletion factors
$(\delta, \kappa_{\rm inv})$ for the 125 GeV Higgs and additional physical
Higgs states.

%%%%%%%%%%%%%%%%%%%%%%%%%%%%%%%%%%%%%%%
%%%%%%%%%%%%%%%%%%%%%%%%%%%%%%%%%%%%%%%
%%%%%%%%%%%%%%%%%%%%%%%%%%%%%%%%%%%%%%%

%\section{Real scalar singlet-Higgs extension(s) \label{sec:scalar_ext}}
\section{Real singlet scalars extension\label{sec:scalar_ext}}
\setcounter{equation}{0}
\setcounter{figure}{0}
\setcounter{table}{0}
\setcounter{footnote}{1}

We now consider a model with the SM particle content and an additional real scalar Higgs $S$ that is a singlet under the SM gauge group with the following scalar
potential\footnote[5]{We treat eq.~(\ref{eq:scalarpotential}) as
an effective scalar potential and therefore the tadpole term
for $S$ that naively seems to follow from eq.~(\ref{eq:scalarpotential})
does not exist in the full theory (see e.g.\ refs.\,\cite{Wells:2002gq,Schabinger:2005ei}).}:
\beq
V(H, S) &=& -m_H^2 H^\dag H + \lambda (H^\dag H)^2 + \frac{1}{2} m_S^2 S^2 + \mu_S H^\dag H S,
\label{eq:scalarpotential}
\eeq
where $H$ is the SM Higgs doublet with a weak hypercharge $Y = 1/2$, and $m_H^2$ is positive such that $H$ acquires a non-zero vacuum expectation value (VEV). The singlet Higgs $S$ is assumed to have invisible decays, governed by the interaction Lagrangian:
\beq
\mathcal{L}_{\rm int} &=& \lambda_\psi S \overline{\psi} \psi,
\label{eq:S_psibar_psi_int}
\eeq
where $\psi \sim ({\bf 1}, {\bf 1}, 0)$ is a very light hidden-sector fermion with a Dirac mass $m_\psi$.
Therefore, the width for $S$ to decay to the invisible states $S \rightarrow \psi \overline \psi$ is 
$\Gamma_{\rm inv} (m_S) \simeq \lambda^2_\psi m_S/8 \pi$,
with the assumption that $m_\psi \ll m_S$.
%Below, we treat $\Gamma_{\rm inv}$ as a free parameter.

At the minimum of the scalar potential, we assume $\langle S \rangle = 0$,
and we use $SU(2)_L$ gauge freedom such that only the neutral component $H^0$ of the SM Higgs doublet acquires a VEV. In unitary gauge, we take
\beq
\langle H^0 \rangle &=& \frac{v + h_{\rm SM}}{\sqrt 2},
% \langle H \rangle = \begin{pmatrix}
% 0\\
% v  + \frac{h_{125}}{\sqrt 2}
% \end{pmatrix},
\eeq
with $v = m_H/\sqrt{\lambda}$ ($\simeq 246$ GeV).
After plugging in for $H$ and expanding around the VEV, eq.~(\ref{eq:scalarpotential}) becomes
\beq
V(h_{\rm SM}, S) &\supset& \frac{1}{2} \begin{pmatrix} h_{\rm SM} & S\end{pmatrix} \mathcal{M}^2 \begin{pmatrix} h_{\rm SM} \\ S \end{pmatrix}
% + \frac{\mu_S}{2} S \tilde{h}^2
% + \mu_S v^2 S
% + \frac{m_H^2}{\sqrt{2} v} \tilde{h}^3
% + \frac{m_H^2}{8 v^2} \tilde{h}^4
,
\eeq
with
\beq
\mathcal{M}^2 &=&
\begin{pmatrix}
2 m_H^2         &   \mu_S v \\
\mu_S v &   m_S^2
\end{pmatrix}.
\eeq
The squared mass matrix can then be diagonalized by a unitary matrix parameterized by a mixing angle
% \beq
% U = \begin{pmatrix}
% \cos \omega & -\sin \omega\\
% \sin \omega & \cos \omega
% \end{pmatrix}
% \eeq
\beq
% \sin 2 \omega = \frac{2 \sqrt{2} \mu_S v}{\sqrt{(m_S^2 - 2 m_H^2)^2 + 8 \mu_S^2 v^2}}, \quad
% \cos 2 \omega = \frac{m_S^2 - 2 m_H^2}{\sqrt{(m_S^2 - 2 m_H^2)^2 + 8 \mu_S^2 v^2}},
\omega &=& \frac{1}{2} \tan^{-1} \frac{2 \mu_S v}{m_S^2 - 2 m_H^2},
\label{eq:mixing_angle}
\eeq
which leads to the physical mass eigenstates $h, \phi$ that are admixtures of the gauge eigenstates $h_{\rm SM}$, $S$:
\beq
h_{\rm SM} &=& \cos \omega \, h + \sin \omega \, \phi,\\
S &=& - \sin \omega \, h + \cos \omega \, \phi,
\eeq
with their corresponding squared masses
\beq
m^2_\pm &=& \frac{1}{2} \left [
(m_S^2 + 2 m_H^2) \pm \sqrt{(m_S^2 - 2 m_H^2)^2 + 4 \mu_S^2 v^2}
\right ]
,
\label{eq:mass_eigenvalues}
\eeq
where $m_+$ is the mass of the heavier state and $m_-$ is the mass of the lighter state.
We take $h$ to be the physical Higgs boson $h$ with mass $m_h \simeq 125$ GeV that was discovered at the LHC in 2012, and the other mass eigenstate $\phi$ to be the exotic Higgs.
From eqs.~(\ref{eq:mixing_angle}) to (\ref{eq:mass_eigenvalues}), note that as $\mu_S$ in eq.~(\ref{eq:scalarpotential}) is tuned down to $0$, the mixing angle $\omega$ vanishes and the mass eigenstates are same as the gauge eigenstates with masses $2 m_H^2$ ($\simeq 125$ GeV) and $m_S^2$.
For purposes of illustration, in this paper, we only consider the possibility of having
$m_\phi$ in the vicinity of $m_h$, specifically
$m_h/2 < m_\phi < 2 m_h$, so that the decays $h \rightarrow \phi \phi$ and
$\phi \rightarrow h h$ are kinematically forbidden.
% Overall, within our assumptions, there are two interesting scenarios:

% \paragraph{$m_h/2 < m_\phi \le m_h$:}
% Here, the already discovered 125 GeV Higgs boson $h$ is the heavier state, and $\phi$ is the lighter state.

% \paragraph{$m_h \le m_\phi < 2 m_h$:}
% Here, the already discovered 125 GeV Higgs boson $h$ is the lighter state, and $\phi$ is the heavier state.

Due to the mixing of the gauge eigenstates $h_{\rm SM}$ and $S$, the physical states $h, \phi$ can now decay to SM states and to invisible states with widths $\Gamma^{h, \phi}_{\rm SM}$ and $\Gamma^{h, \phi}_{\rm inv}$ respectively.
Apart from the masses $m_\pm$ (of which one is already known to be $\sim$125 GeV), the widths, the production cross-sections and branching fractions of ${h, \phi}$ can be expressed in terms of only two free parameters $\delta$ and $\kappa_{\rm inv}$ that are defined as:
\beq
\delta &\equiv& \left|\sin \omega\right|,\\
\kappa_{\rm inv} &\equiv& \Gamma_{\rm inv}/\Gamma^{125}_{\rm SM}
\eeq
In terms of $\Gamma_{\rm SM}$ (total width for $h_{\rm SM}$ to SM states) and $\Gamma_{\rm inv}$ (width for $S \rightarrow \psi \overline \psi$):
\begin{alignat}{2}
\Gamma^\phi_{\rm SM} &= \delta^2 \ \Gamma_{\rm SM} (m_\phi),
\qquad
&&\Gamma^\phi_{\rm inv} = (1 - \delta^2) \ \Gamma_{\rm inv},
\\
\Gamma^h_{\rm SM} &= (1 - \delta^2) \ \Gamma^{125}_{\rm SM},
\qquad
&&\Gamma^h_{\rm inv} = \delta^2 \ \Gamma_{\rm inv},
\end{alignat}
where $\Gamma_{\rm inv}$ is treated as a free parameter,
and $\Gamma^{125}_{\rm SM} = \Gamma_{\rm SM}(\text{125 GeV})$.
Also, in terms of the production cross-section $\sigma_{\rm SM}$ of $h_{\rm SM}$, the production cross-sections for ${h, \phi}$ are
\beq
\sigma^\phi &=& \delta^2 \, \sigma_{\rm SM} (m_\phi),\\
\sigma^h &=& (1 - \delta^2) \, \sigma^{125}_{\rm SM}.
\label{eq:Singlet_sigmah}
\eeq

We can now compare the production cross-sections times branching ratios of ${h, \phi}$ to the ones in the SM.
In order to do so, we first note that the total widths $\Gamma^{h, \phi}$ of ${h, \phi}$ are related to the SM width $\Gamma_{\rm SM}$ via:
\beq
\frac{\Gamma^\phi}{\Gamma_{\rm SM}} &=&
\frac{\kappa_{\rm inv}
+
\left(
\kappa_{\rm SM} - \kappa_{\rm inv}
\right) \delta^2}{\kappa_{\rm SM}},\\
\frac{\Gamma^h}{\Gamma^{125}_{\rm SM}} &=&  1 - (1 - \kappa_{\rm inv})\delta^2,
\label{eq:Singlet_Gammah}
\eeq
where we have defined
\beq
\kappa_{\rm SM} &\equiv& \Gamma_{\rm SM} (m_\phi)/\Gamma^{125}_{\rm SM}.
\eeq
Using these results and definitions, 
the ratio of branching ratios of ${h, \phi}$ into $j^{\rm th}$ SM final state $B^{h, \phi}_j = \Gamma^{h, \phi}_j/\Gamma^{h, \phi}$ to that of the SM $B_{{\rm SM}, j}$ can be expressed as
\beq
\frac{B^\phi_j}{B_{{\rm SM}, j}} &=&
\frac{\delta^2 \, \kappa_{\rm SM}}
{\kappa_{\rm inv}
+
\left(
\kappa_{\rm SM} - \kappa_{\rm inv}
\right)
\delta^2},\\
\frac{B^h_j}{B^{125}_{{\rm SM}, j}} &=& \frac{1 - \delta^2}{1 - (1 - \kappa_{\rm inv})\delta^2},
\label{eq:Singlet_Bh}
\eeq
so the production rates of ${h, \phi}$ in $j^{\rm th}$ SM final state is
\beq
\frac{\sigma^\phi B^\phi_j}{\sigma_{\rm SM} B_{{\rm SM}, j}} &=& \frac{\delta^4 \, \kappa_{\rm SM}}
{\kappa_{\rm inv}
+
\left(
\kappa_{\rm SM} - \kappa_{\rm inv}
\right)
\delta^2},\label{eq:sigmaBR_phi}\\
\frac{\sigma^h B^h_j}{\sigma^{125}_{\rm SM} B^{125}_{{\rm SM}, j}} &=& \frac{(1 - \delta^2)^2}{1 - (1 - \kappa_{\rm inv})\delta^2}.
\eeq
Finally, the invisible branching ratios $B^{h, \phi}_{\rm inv} = \Gamma^{h, \phi}_{\rm inv}/\Gamma^{h, \phi}$ of ${h, \phi}$:
\beq
B^\phi_{\rm inv} &=& \frac{(1 - \delta^2) \, \kappa_{\rm inv}}{\kappa_{\rm inv} + (\kappa_{\rm SM} - \kappa_{\rm inv})\delta^2},\\
B^h_{\rm inv} &=& \frac{\delta^2 \, \kappa_{\rm inv}}{1 - (1 - \kappa_{\rm inv}) \delta^2}.
\label{eq:Singlet_Binvh}
\eeq
Note that the expressions for $h$ in eqs.~(\ref{eq:Singlet_sigmah}),
(\ref{eq:Singlet_Gammah}), (\ref{eq:Singlet_Bh}), and
(\ref{eq:Singlet_Binvh}) match with eqs.~(\ref{eq:sigma}),
(\ref{eq:SM_Gammah}), (\ref{eq:BSMj}), and
(\ref{eq:Binv}) respectively.

Using the above equations, we can now obtain the
current constraints/future sensitivities for real singlet
scalar extension to the SM that impose the depletion factors $(\delta, \kappa_{\rm inv})$
on the 125 GeV Higgs boson $h$ and also predict an exotic Higgs $\phi$.
The indirect constraints and projections from the precision probes for the
125 GeV Higgs boson $h$ were obtained in the previous section (Section~\ref{sec:SM_Higgs_bounds}).
Whereas, the current constraints are considered in the next section (Section~\ref{sec:Singlet_Higgs_bounds})
along with some future projections for $(\delta, \kappa_{\rm inv})$
from the direct searches for the exotic Higgs boson $\delta$ with $m_h/2 \le m_\phi \le 2 m_h$
(for simplicity and easy compatibility with precision electroweak constraints).
And, we will see that the precision probes for the observed Higgs boson $h$ are typically much more
constraining than the direct searches for the exotic Higgs $\phi$.

%%%%%%%%%%%%%%%%%%%%%%%%%%%%%%%%%%%%%%%
%%%%%%%%%%%%%%%%%%%%%%%%%%%%%%%%%%%%%%%
%%%%%%%%%%%%%%%%%%%%%%%%%%%%%%%%%%%%%%%
\section{Constraints on real singlet scalars\label{sec:Singlet_Higgs_bounds}}
%\subsection{Constraints on ${(\delta, \kappa_{\rm inv}, m_\phi)}$ \label{sec:Singlet_Higgs_bounds}}
% \setcounter{equation}{0}
% \setcounter{figure}{0}
% \setcounter{table}{0}
% \setcounter{footnote}{1}
\setcounter{equation}{0}
\setcounter{figure}{0}
\setcounter{table}{0}
\setcounter{footnote}{1}

As mentioned above, in the present case the sine-squared of the mixing angle between the gauge eigenstates $h_{\rm SM}$ and $S$ gives rise to the universal depletion factor $\delta$ for the couplings of the
physical 125 GeV Higgs boson $h$, and the invisible width $\Gamma_{\rm inv}$ of $S$
gives rise to the invisible width of $h$.
Therefore, all of the constraints on $(\delta, \kappa_{\rm inv})$ from the precision probes for
the 125 GeV Higgs boson, considered in Section~\ref{sec:125GeVHiggs}, are directly applicable
for $(\delta, \kappa_{\rm inv})$ in the case of real scalar singlet extension to the SM.

Apart from the indirect constraints that come from the precision probes for the 125 GeV Higgs boson, there are also
additional bounds on the exotic Higgs $\phi$ for various values of its mass $m_\phi$ from precision
electroweak observables,
and also from the collider searches
for an additional neutral Higgs boson over a wide range of masses.

First, let us consider the bounds from precision electroweak observables, in particular from
the Peskin-Takeuchi parameters $S, T$, and $U$ \cite{Peskin:1991sw}.
To that end we begin by noting the one-loop contribution of the SM Higgs with mass $m$ to
the massive vector boson $(V = W, Z)$ propagators:
\beq
\Pi_{\rm V V} (p^2; m)
&=&
- \frac{\alpha_e}{4 \pi} \frac{m_V^2}{s_W^2 m_W^2}
\left[
\frac{1}{2} A_0 (m) +
m_V^2 B_0(p^2; m_V, m) - B_{00}(p^2; m_V, m)
\right],
%\notag\\
\label{eq:vacuum_pol_WZ}
\eeq
where $\alpha_e$ is the fine structure constant, $s_W$ ($c_W$) is the sine (cosine) of the
weak-mixing angle, and $A_0 (m_0)$, $B_0(p^2; m_1, m_0)$, and $B_{00}(p^2; m_1, m_0)$ are
the Passarino-Veltman functions following the conventions of ref.\,\cite{Patel:2015tea}.
Then the predictions for the Peskin-Takeuchi parameters in the real singlet scalar extension
to the SM are given by:
\beq
S &=& \frac{4 c_W^2 s_W^2}{\alpha_e} \left[ \frac{\Pi^{\rm new}_{ZZ} (m_Z^2) - \Pi^{\rm new}_{ZZ} (0)}{m_Z^2}\right],\\
T &=& \frac{1}{\alpha_e} \left[ \frac{\Pi^{\rm new}_{WW} (0)}{m_W^2} - \frac{\Pi^{\rm new}_{ZZ} (0)}{m_Z^2}\right],\\
U &=& \frac{4 s_W^2}{\alpha_e} \left[ \frac{\Pi^{\rm new}_{WW}(m_W^2) - \Pi^{\rm new}_{WW}(0)}{m_W^2} \right] - S,
\eeq
with
\beq
\Pi^{\rm new}_{VV} (p^2) = (1 - \delta^2) \ \Pi_{VV} (p^2; m_h) + \delta^2 \ \Pi_{VV} (p^2; m_\phi) - \Pi_{VV} (p^2; m_h),
\eeq
such that the parameters $S$, $T$, and $U$ account for the contribution of the physical Higgs
boson with mass $m_h$ that is already included in them, and therefore are normalized to
reflect only the new physics (i.e real scalar singlet Higgs) contribution.

From the global electroweak fit at NNLO by the Gfitter group \cite{Baak:2014ora},
the current experimental measurements of the Peskin-Takeuchi parameters,
with the SM reference point taken to be $(m_h, m_t) = (125, 173)$ GeV, are
\beq
\hat S = 0.05, \quad \hat T = 0.09, \quad \hat U = 0.01,
\eeq
with the corresponding uncertainties ($\sigma_i$)
\beq
\sigma_S = 0.11, \quad \sigma_T = 0.13, \quad \sigma_U = 0.11,
\eeq
and the correlation coefficients ($\rho_{i j} = \rho_{j i} = \frac{\sigma^2_{i j}}{\sigma_i \sigma_j}$)
\beq
\rho_{S T} = 0.90, \quad \rho_{S U} = -0.59, \quad \rho_{T U} = -0.83.
\eeq
In order to finally obtain a bound on $\delta$, we check the compatibility of the model
predictions with that of the experimental measurements.
In particular, for a chosen $m_\phi$, we compute the $\chi^2$-value,
which is a function of $\delta$,
using
\beq
\chi^2 = \bold{y}_i \,\bold{V}^{-1}_{i j} \,\bold{y}_j,
\eeq
with $\bold{y}_i = (S - \hat S, T - \hat T, U - \hat U)$, and the covariance matrix elements
$\bold{V}_{i i} = \sigma_i^2$ and $\bold{V}_{i j} = \rho_{i j} \sigma_i \sigma_j$ ($i \neq j$).
And, solve for $\delta$ after requiring $\chi^2 = 7.81$ corresponding to
a fit with 
95\% confidence level (or a $p$-value of 0.05) with 3 degrees of freedom \cite{ParticleDataGroup:2020ssz}. 

%%%%%%%%%%%%%%%%%%%%%%%%%%%%%%%%%%%%%%%%%%%%%%%%%%%%%%%%%%%%%%%%%%%%%%%%%%%%%%%%%%
\begin{figure}
  \begin{center}\begin{minipage}[]{0.95\linewidth}
    \includegraphics[width=15cm]{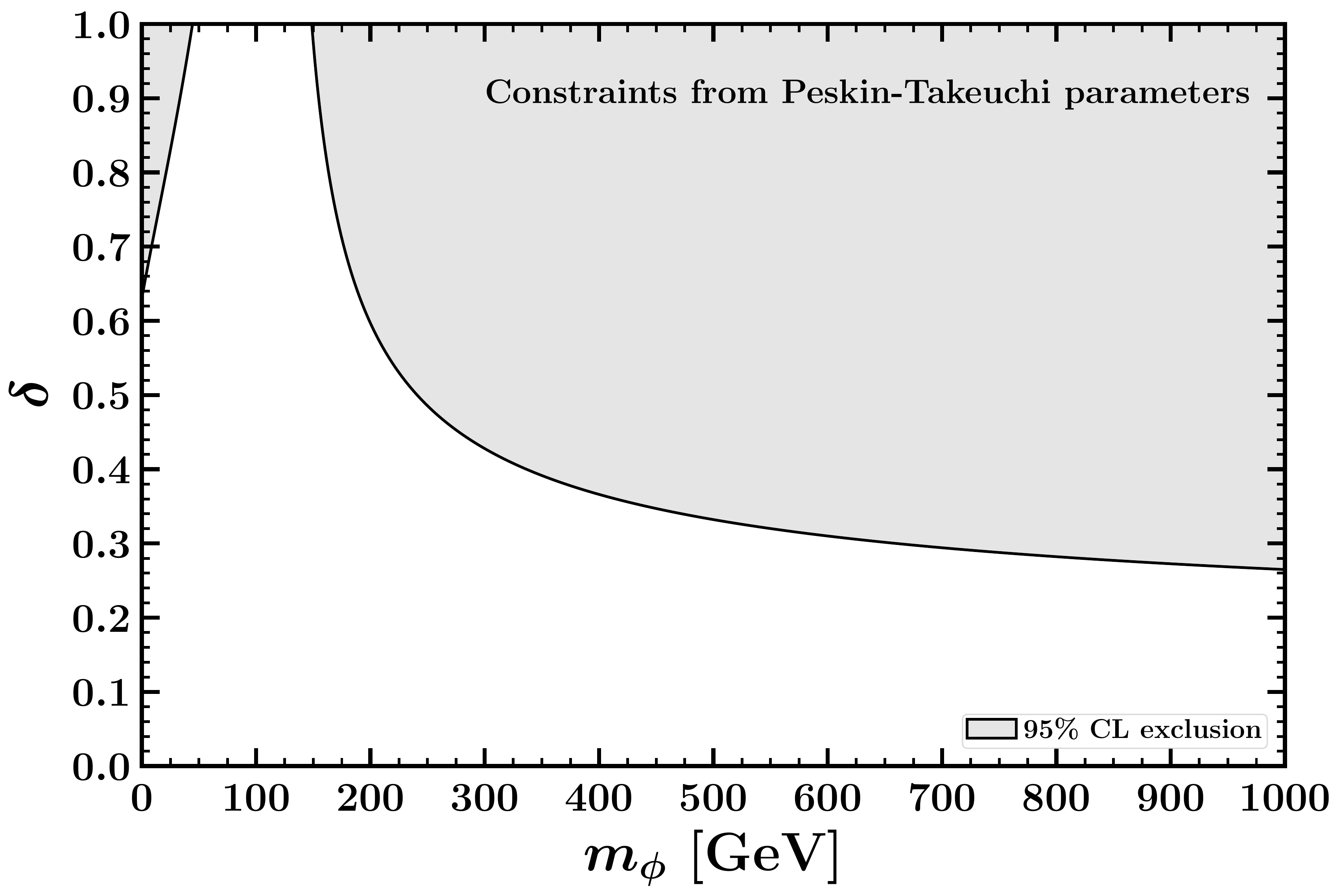}
  \end{minipage}\end{center}
 \caption{Constraints on $\delta$ as a function of the exotic Higgs mass $m_\phi$ from precision electroweak observables $S$, $T$, and $U$ parameters. The bounds are extracted by evaluating the $\chi^2$ using the model predictions and observed values of Peskin-Takeuchi
 parameters, taking all the experimental uncertainties and correlations into account,
 and demanding that the resulting $\chi^2$ corresponds to a 95\% confidence level fit.
\label{fig:PEW_bounds}}
\end{figure}
%%%%%%%%%%%%%%%%%%%%%%%%%%%%%%%%%%%%%%%%%%%%%%%%%%%%%%%%%%%%%%%%%%%%%%%%%%%%%%%%%% 
Figure~\ref{fig:PEW_bounds} shows the bound obtained from Peskin-Takeuchi parameters using
the procedure detailed above on $\delta$ as a function of $m_\phi$.
Actually, for a fixed significance level, we find a disagreement in comparing our results with the ones obtained in refs.\,\cite{Robens:2016xkb,Robens:2015gla}. The reason for this disagreement is due to an
additional factor of $-1$ that is included in the vacuum polarization function $\Pi_{V V}$
which can be inferred from (eqs.~(24)~and~(25) of) ref.\,\cite{Lopez-Val:2014jva}\footnote[4]{Additionally, in eqs.~(24)~and~(25) of ref.\,\cite{Lopez-Val:2014jva}, we note that there is an additional factor of $\frac{1}{2}$ in the term with $A_0(m)$, but that term drops out of the Peskin-Takeuchi paramaters as it is independent of $p^2$.}
which seemed to have propagated into the results in refs.\,\cite{Robens:2016xkb,Robens:2015gla}.
Taking $\Pi_{V V} \rightarrow -\Pi_{V V}$ in eq.~(\ref{eq:vacuum_pol_WZ}), we exactly
reproduce the bounds from $S, T,$ and $U$ parameters reported in refs.\,\cite{Robens:2016xkb,Robens:2015gla}.
To check the correctness of the sign in eq.~(\ref{eq:vacuum_pol_WZ}), we have computed the one-loop
contribution of the SM Higgs to $s_W^2$ and checked that it agrees with the results in
the standard quantum field theory textbooks e.g.\ refs.\,\cite{Peskin:1995ev,Schwartz:2014sze}.
Furthermore, the constraints we obtain from $S$, $T$, and $U$ parameters are in
a good agreement with that of ref.\,\cite{Dawson:2021jcl} which uses the experimental inputs
from ref.\,\cite{ParticleDataGroup:2020ssz}.
We note that our results do not change significantly if we use experimental inputs from ref.\,\cite{ParticleDataGroup:2020ssz} instead of ref.\,\cite{Baak:2014ora}.
Therefore, in contrast to the claim in refs.\,\cite{Robens:2016xkb,Robens:2015gla}, $S$, $T$, and $U$ electroweak
precision parameters pose significant constraints on $\delta$ as a function of $m_\phi$, especially
for large $m_\phi$.

%%%%%%%%%%%%%%%%%%%%%%%%%%%%%%%%%%%%%%%%%%%%%%%%%%%%%%%%%%%%%%%%%%%%%%%%%%%%%%%%%%%%%%%%%%%
\begin{table}
\begin{minipage}[]{0.95\linewidth}
\caption{Collider searches for the decays of (additional) neutral Higgs boson(s) to (visible) SM and invisible final states over wide range of masses.
For convenience, various searches are assigned labels which will be used in the figures below.
$B_{{\rm SM}, j}$ below is the SM branching ratio for $h_{\rm SM} \rightarrow j^{\rm th}$ SM final state.
\label{tab:phi_bounds}}
\end{minipage}
\begin{center}
\begin{tabular}{|c | c | c | c |}
\hline
~Label~ & ~Mass range (GeV)~  &  ~Limit reported (95\% CL)~  & ~Reference~\\
\hline
\hline
~LEP 2005 ($\phi \rightarrow \text{inv}$)~ 
&
~$50-110$~ 
&  
~$\mu_{\rm inv}$~
&
~\cite{L3:2004svb}~
\\[1pt]
\hline
~LEP 2011 ($\phi \rightarrow \text{inv}$)~ 
&
~$90-118$~ 
&  
~$\mu_{\rm inv}$~
&
~\cite{LEPHiggsWorkingforHiggsbosonsearches:2001ypa}~
\\[1pt]
\hline
~CMS 2018 ($\phi \rightarrow \text{inv}$)~ 
&
~$110-1000$~ 
&  
~$\mu_{\rm inv}$~
&
~\cite{CMS:2018yfx}~
\\[1pt]
\hline
~ATLAS 2019 ($\phi \rightarrow \text{inv}$)~ 
&
~$75-3000$~ 
&  
~$\mu_{\rm inv} \, \sigma_{\rm SM}$~
&
~\cite{ATLAS:2018bnv}~
\\[1pt]
\hline
~ATLAS 2022 ($\phi \rightarrow \text{inv}$)~ 
&
~$50-2000$~ 
&  
~$\mu_{\rm inv} \, \sigma_{\rm SM}$~
&
~\cite{ATLAS:2022yvh}~
\\[1pt]
\hline
~LEP 2006 ($\phi \rightarrow b b$)~ 
&
~$12-120$~ 
&  
~$\mu \, B_{{\rm SM}, b b}$~
&
~\cite{ALEPH:2006tnd}~
\\[1pt]
\hline
~CMS 2012 (combined)~ 
&
~$110-1000$~ 
&  
~$\mu$~
&
~\cite{CMS:2012qwq}~
\\[1pt]
\hline
~CMS 2013 ($\phi \rightarrow W W$)~ 
&
~$110-600$~ 
&  
~$\mu$~
&
~\cite{CMS:2013cul}~
\\[1pt]
\hline
~CMS 2015 ($\phi \rightarrow WW/ZZ$)~ 
&
~$145-1000$~ 
&  
~$\mu$~
&
~\cite{CMS:2015hra}~
\\[1pt]
\hline
~ATLAS 2015 ($\phi \rightarrow ZZ$)~ 
&
~$140-1000$~ 
&  
~$\mu \, \sigma_{\rm SM} \, B_{{\rm SM}, Z Z}$~
&
~\cite{ATLAS:2015pre}~
\\[1pt]
\hline
~ATLAS 2020 ($\phi \rightarrow ZZ$)~ 
&
~$210-2000$~ 
&  
~$\mu \, \sigma_{\rm SM} \, B_{{\rm SM}, Z Z}$~
&
~\cite{ATLAS:2020tlo}~
\\[1pt]
\hline
~CMS 2018 ($\phi \rightarrow ZZ$)~ 
&
~$130-3000$~ 
&  
~$\mu \, \sigma_{\rm SM} \, B_{{\rm SM}, Z Z}$~
&
~\cite{CMS:2018amk}~
\\[1pt]
\hline
\end{tabular}
\end{center}
\end{table}
%%%%%%%%%%%%%%%%%%%%%%%%%%%%%%%%%%%%%%%%%%%%%%%%%%%%%%%%%%%%%%%%%%%%%%%%%%%%%%%%%%%%%%%%%%%
%%%%%%%%%%%%%%%%%%%%%%%%%%%%%%%%%%%%%%%%%%%%%%%%%%%%%%%%%%%%%%%%%%%%%%%%%%%%%%%%%%
\begin{figure}
  \begin{minipage}[]{0.495\linewidth}
    \includegraphics[width=8cm]{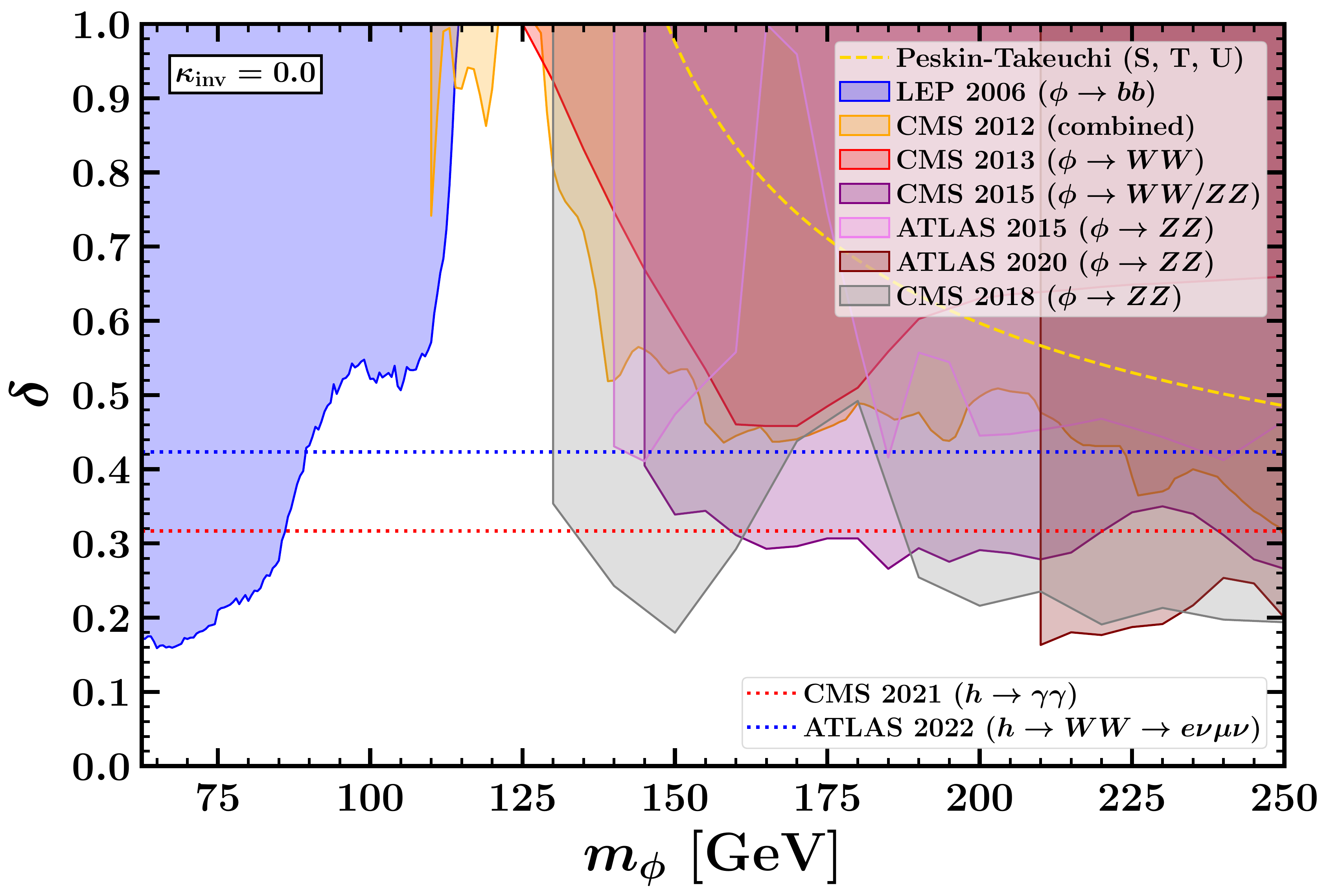}
  \end{minipage}
  \begin{minipage}[]{0.495\linewidth}
    \includegraphics[width=8cm]{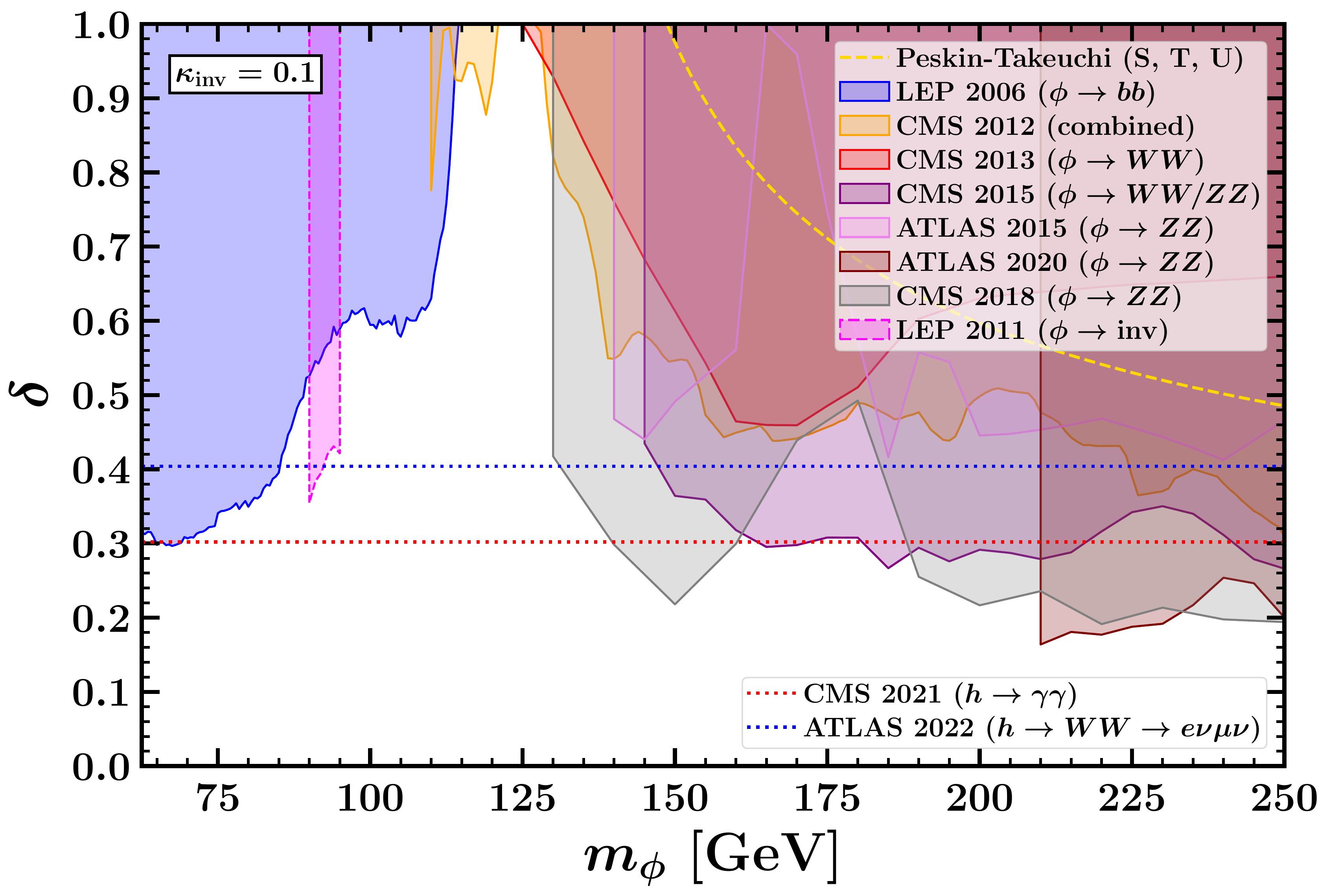}
  \end{minipage}
  \vspace{0.2cm}
  \begin{minipage}[]{0.495\linewidth}
    \includegraphics[width=8cm]{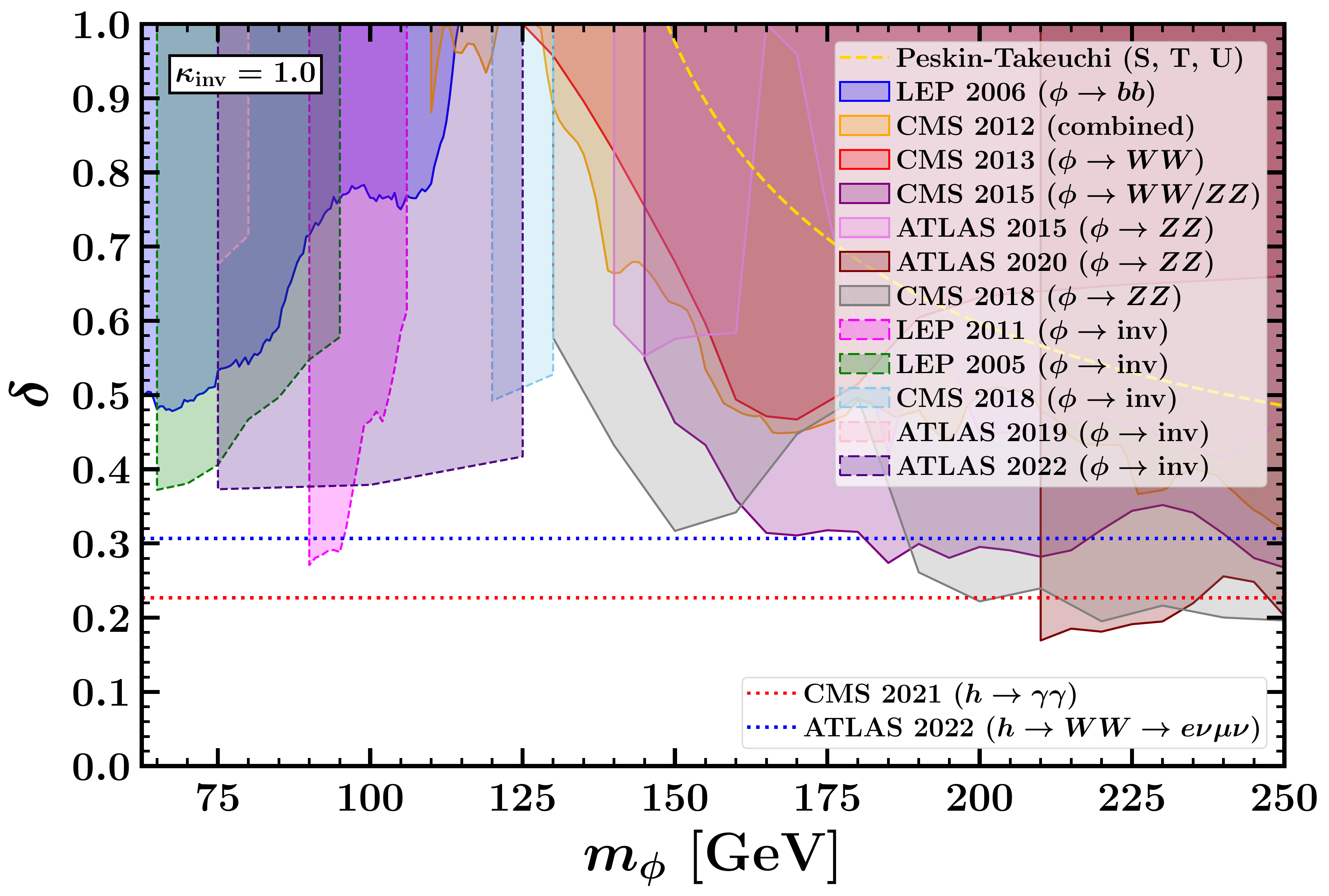}
  \end{minipage}
  \begin{minipage}[]{0.495\linewidth}
    \includegraphics[width=8cm]{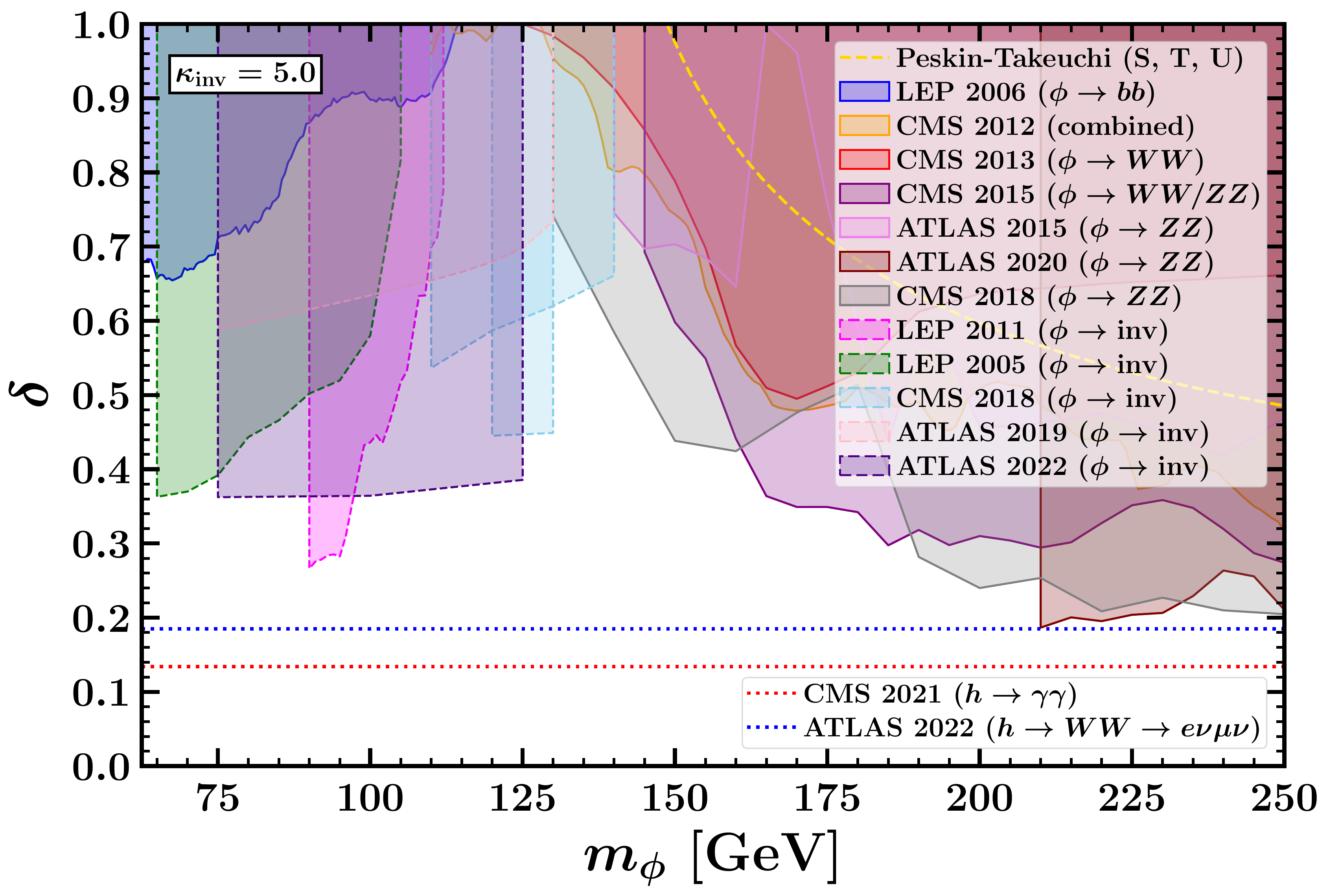}
  \end{minipage}
  \vspace{0.2cm}
  \begin{minipage}[]{0.495\linewidth}
    \includegraphics[width=8cm]{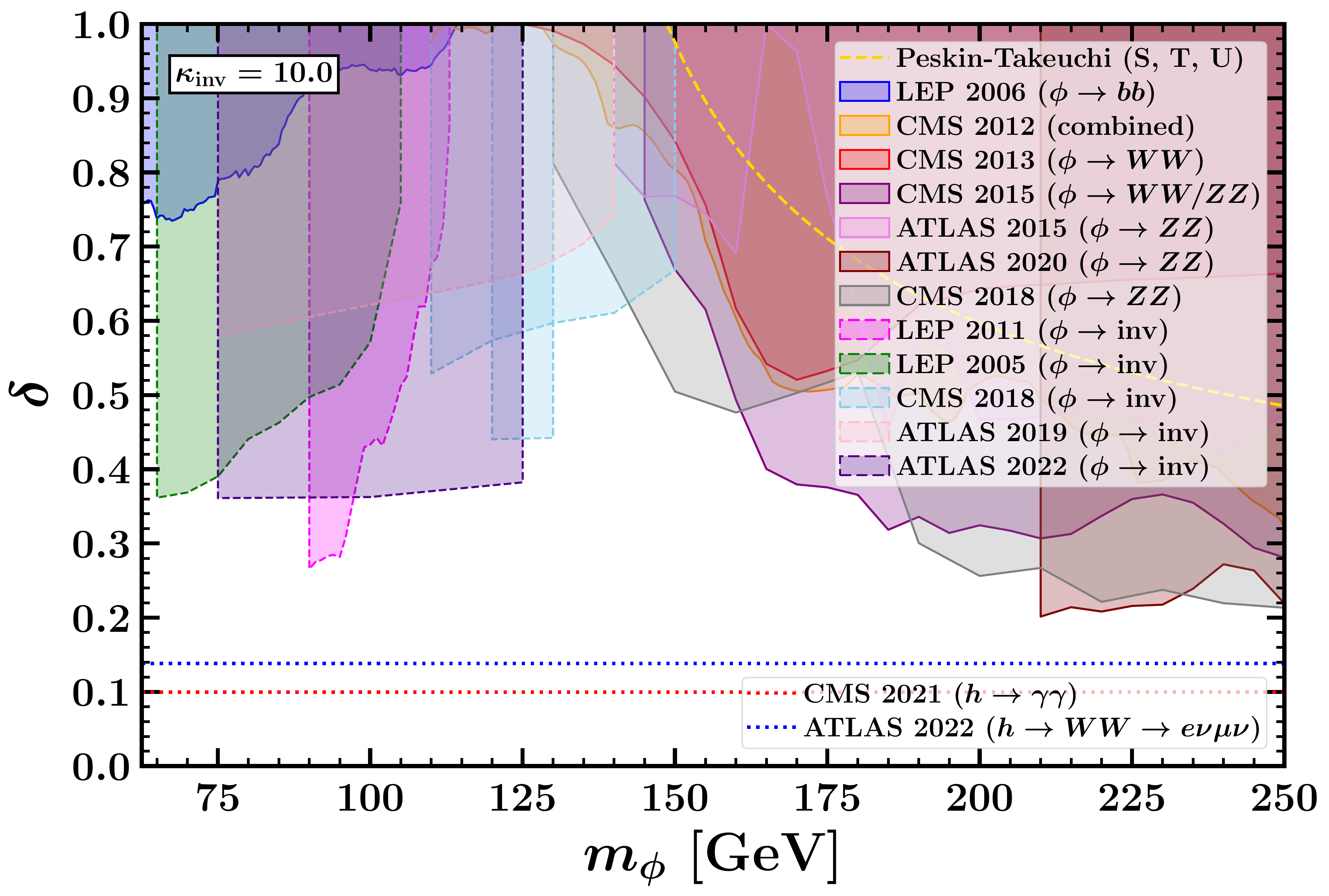}
  \end{minipage}
  \begin{minipage}[]{0.495\linewidth}
    \includegraphics[width=8cm]{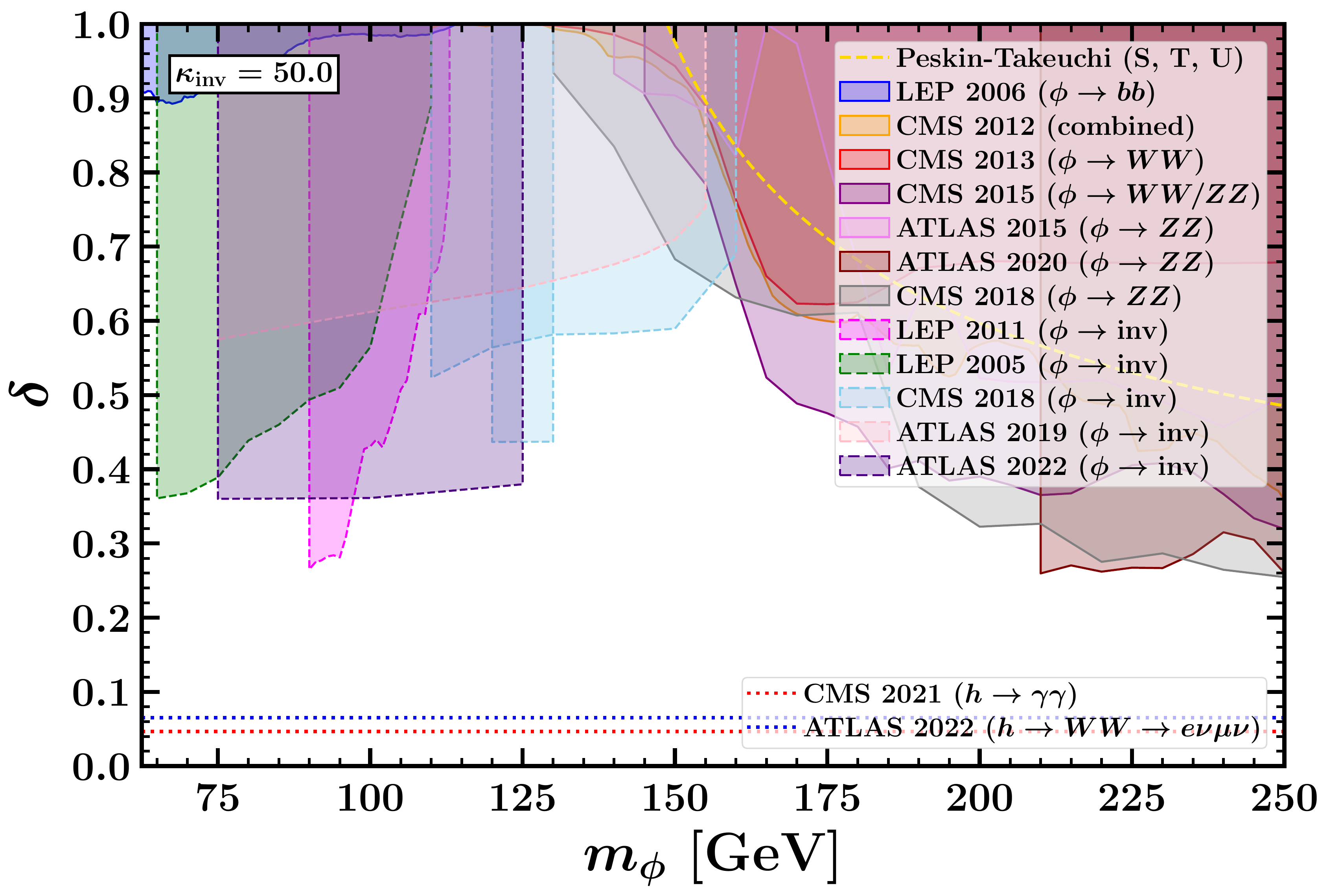}
  \end{minipage}
 \caption{\label{fig:phi_bounds}95\% CL bounds on $\delta$ as a function of the exotic Higgs mass $m_\phi$
 for various choices of $\kappa_{\rm inv} = 0, 0.1, 1, 5, 10, 50$. The shaded regions with
 solid borders are the bounds from the searches for $\phi \rightarrow$ (visible) SM final states,
 and the shaded regions with dashed borders are the bounds from the invisible searches for
 $\phi$. The labels for each of these shaded regions correspond to the ones listed in Table~\ref{tab:phi_bounds}.
 Gold dashed lines show the constraints from electroweak precision observables,
 namely the Peskin-Takeuchi $S$, $T$, and $U$ parameters. The dotted lines show the indirect constraints from the precision probes for $h \rightarrow W W \rightarrow e \nu \mu \nu$ and $h \rightarrow \gamma \gamma$ (strongest bound in most cases).}
\end{figure}
%%%%%%%%%%%%%%%%%%%%%%%%%%%%%%%%%%%%%%%%%%%%%%%%%%%%%%%%%%%%%%%%%%%%%%%%%%%%%%%%%%
%%%%%%%%%%%%%%%%%%%%%%%%%%%%%%%%%%%%%%%%%%%%%%%%%%%%%%%%%%%%%%%%%%%%%%%%%%%%%%%%%%
\begin{figure}
  \begin{minipage}[]{0.495\linewidth}
    \includegraphics[width=8cm]{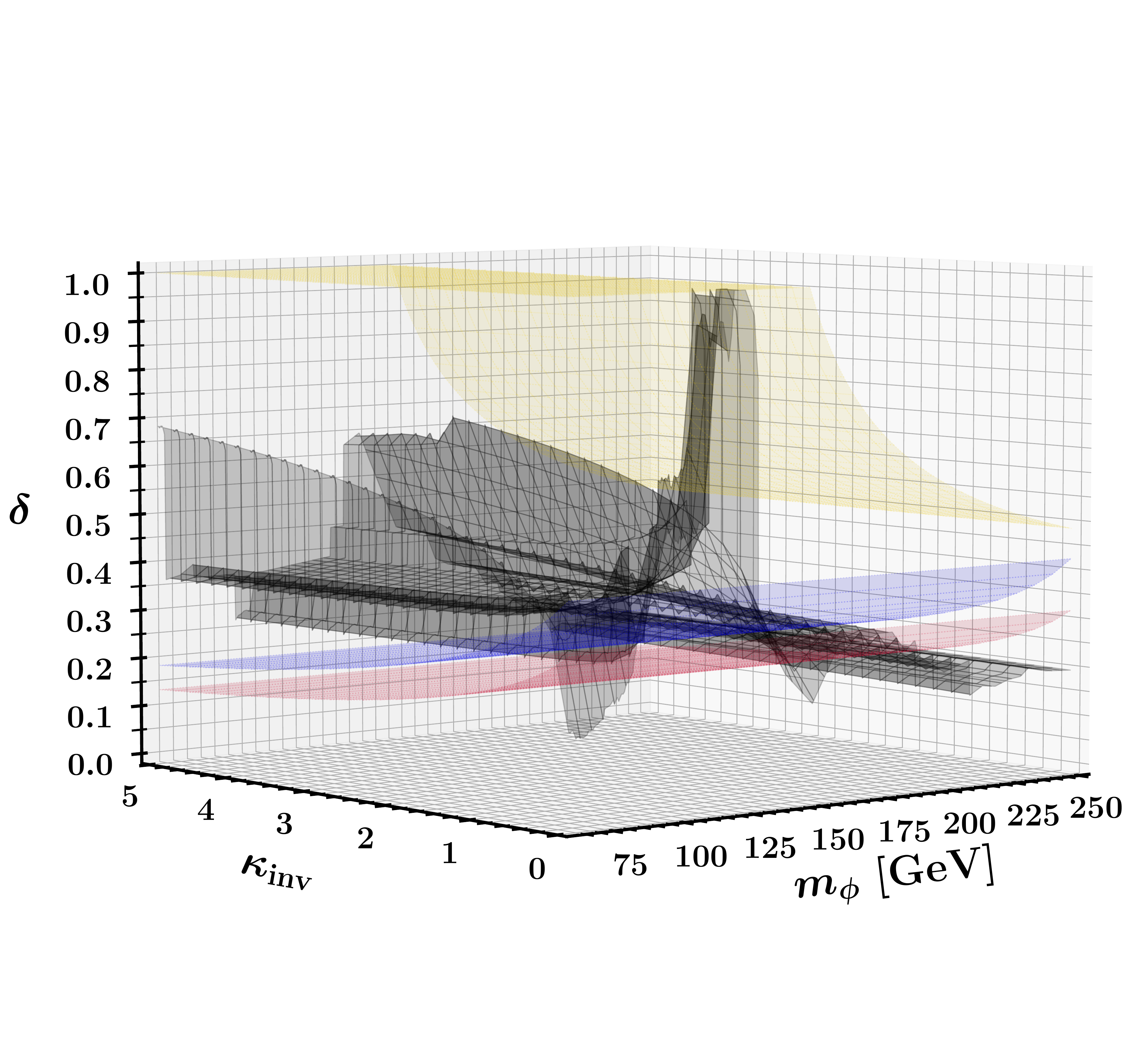}
  \end{minipage}
  \begin{minipage}[]{0.495\linewidth}
    \includegraphics[width=8cm]{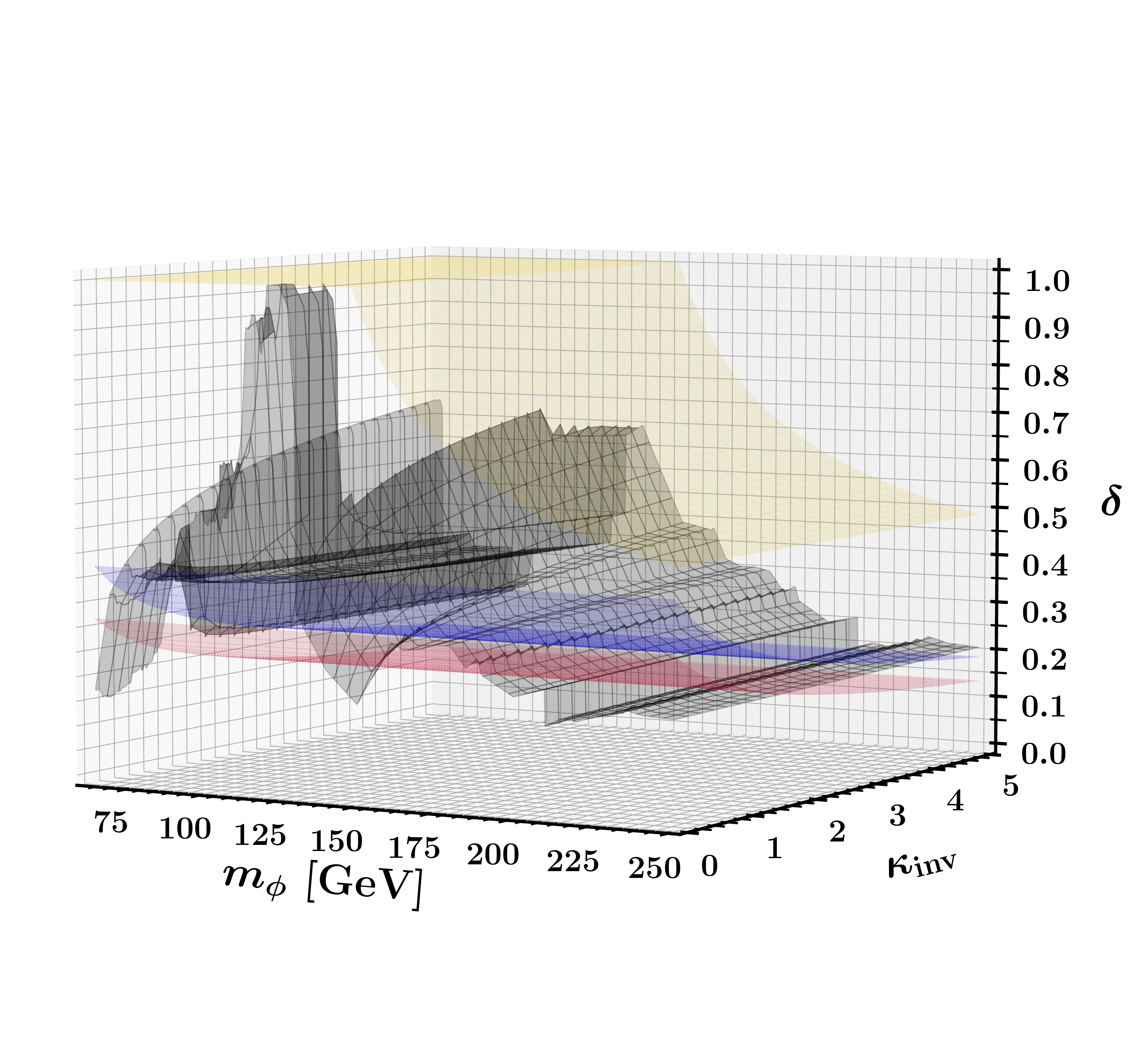}
  \end{minipage}
 \caption{\label{fig:phi_bounds_3D}Constraints on the 3D parameter space $(\delta, \kappa_{\rm inv}, m_\phi)$ from the collider searches for $\phi$ (black shaded envelope; listed in Table~\ref{tab:phi_bounds}), along with the strongest constraints from the precision probes for $h \rightarrow \gamma \gamma$ (red) and $h \rightarrow WW \rightarrow e \nu \mu \nu$ (blue). Also shown are the constraints from the Peskin-Takeuchi parameters for the SM extended with a real SM gauge-singlet scalar Higgs (gold).
 The 3D plots here showcase various bounds on all the free parameters of this model, and more details are shown in Figure~\ref{fig:phi_bounds} with some 2D slices with fixed $\kappa_{\rm inv}$.}
\end{figure}
%%%%%%%%%%%%%%%%%%%%%%%%%%%%%%%%%%%%%%%%%%%%%%%%%%%%%%%%%%%%%%%%%%%%%%%%%%%%%%%%%%
We now turn to constraints on our parameter space from the collider searches for
additional neutral Higgs boson.
Once again, for invisible searches, an upper bound is given on $\mu_{\rm inv}$ (defined in
eq.~(\ref{eq:mu_inv})) which translates into
\beq
\mu_{\rm inv} = \frac{\delta^2 (1 - \delta^2) \, \kappa_{\rm inv}}{\kappa_{\rm inv} + (\kappa_{\rm SM} - \kappa_{\rm inv})\delta^2},
\eeq
for the exotic Higgs $\phi$. For a bound reported on $\mu_{\rm inv}$ at each $m_\phi$ and for a fixed
$\kappa_{\rm inv}$, we can then solve for the allowed values of $\delta$ using the
above equation.

For searches for additional Higgs boson in $j^{\rm th}$ SM final state, an upper limit is reported on
\beq
\mu
&=&
\frac{\sigma \, B_j}{\sigma_{\rm SM} \, B_{{\rm SM}, j}}.
\eeq
For $\phi$ boson, this limit can be recasted, using eq.~(\ref{eq:sigmaBR_phi}), as
\beq
\mu &=& \frac{\delta^4 \, \kappa_{\rm SM}}
{\kappa_{\rm inv}
+
\left(
\kappa_{\rm SM} - \kappa_{\rm inv}
\right)
\delta^2},
\eeq
from which we can extract the allowed values of $\delta$ for a fixed $\kappa_{\rm inv}$
for each $m_\phi$. 
Various collider searches that look for (additional) neutral Higgs boson(s) in various decay channels (including invisible searches) that pose significant constraints on the parameter space of $(\delta, \kappa_{\rm inv}, m_\phi)$ are listed in Table~\ref{tab:phi_bounds}. The relevant data associated with these searches was obtained, in part, from the files provided with the publicly available Fortran code {\sc HiggsBounds5} \cite{Bechtle:2020pkv}.
And the widths and branching ratios of the SM Higgs boson for various masses are obtained using
the program {\sc HDECAY} \cite{Djouadi:2018xqq,Djouadi:1997yw}.

In Figure~\ref{fig:phi_bounds} we show various constraints on $\delta$ as a function of
the exotic Higgs mass $m_\phi$ for various choices of $\kappa_{\rm inv}$ at 95\% CL.
The shaded regions with solid borders are the bounds from the searches for the exotic Higgs $\phi$
decaying to SM final states,
and the shaded regions with dashed borders are the bounds from the invisible searches for
$\phi$.
The labels for each of these shaded regions correspond to the ones listed in Table~\ref{tab:phi_bounds}.
Gold dashed lines show the constraints from electroweak precision observables,
namely the Peskin-Takeuchi $S$, $T$, and $U$ parameters.
And, dotted lines show the bounds from the precision probes for $h \rightarrow W W \rightarrow e \nu \mu \nu$ and $h \rightarrow \gamma \gamma$ which typically are much more constraining than the searches for $\phi$.
The 3D plots in Figure~\ref{fig:phi_bounds_3D} show the richness of various bounds from the searches for neutral Higgs bosons in various decay channels on the parameter space of $(\delta, \kappa_{\rm inv}, m_\phi)$. Various 2D slices of the 3D plots in Figure~\ref{fig:phi_bounds_3D} with fixed $\kappa_{\rm inv}$ were shown in Figure~\ref{fig:phi_bounds}.

%%%%%%%%%%%%%%%%%%%%%%%%%%%%%%%%%%%%%%%%%%%%%%%%%%%%%%%%%%%%%%%%%%%%%%%%%%%%%%%%%%
\begin{figure}
  \begin{minipage}[]{0.495\linewidth}
    \includegraphics[width=8cm]{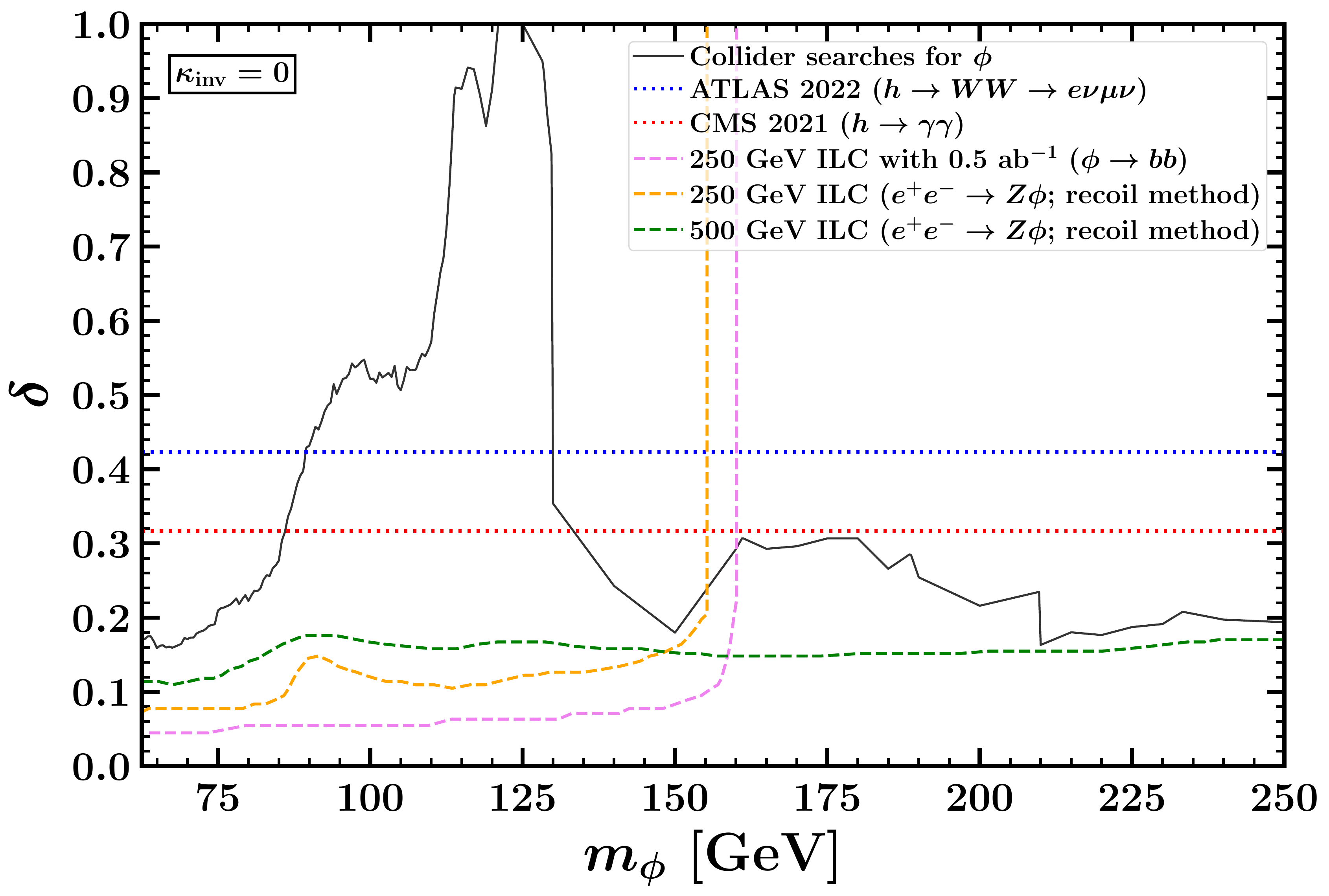}
  \end{minipage}
  \begin{minipage}[]{0.495\linewidth}
    \includegraphics[width=8cm]{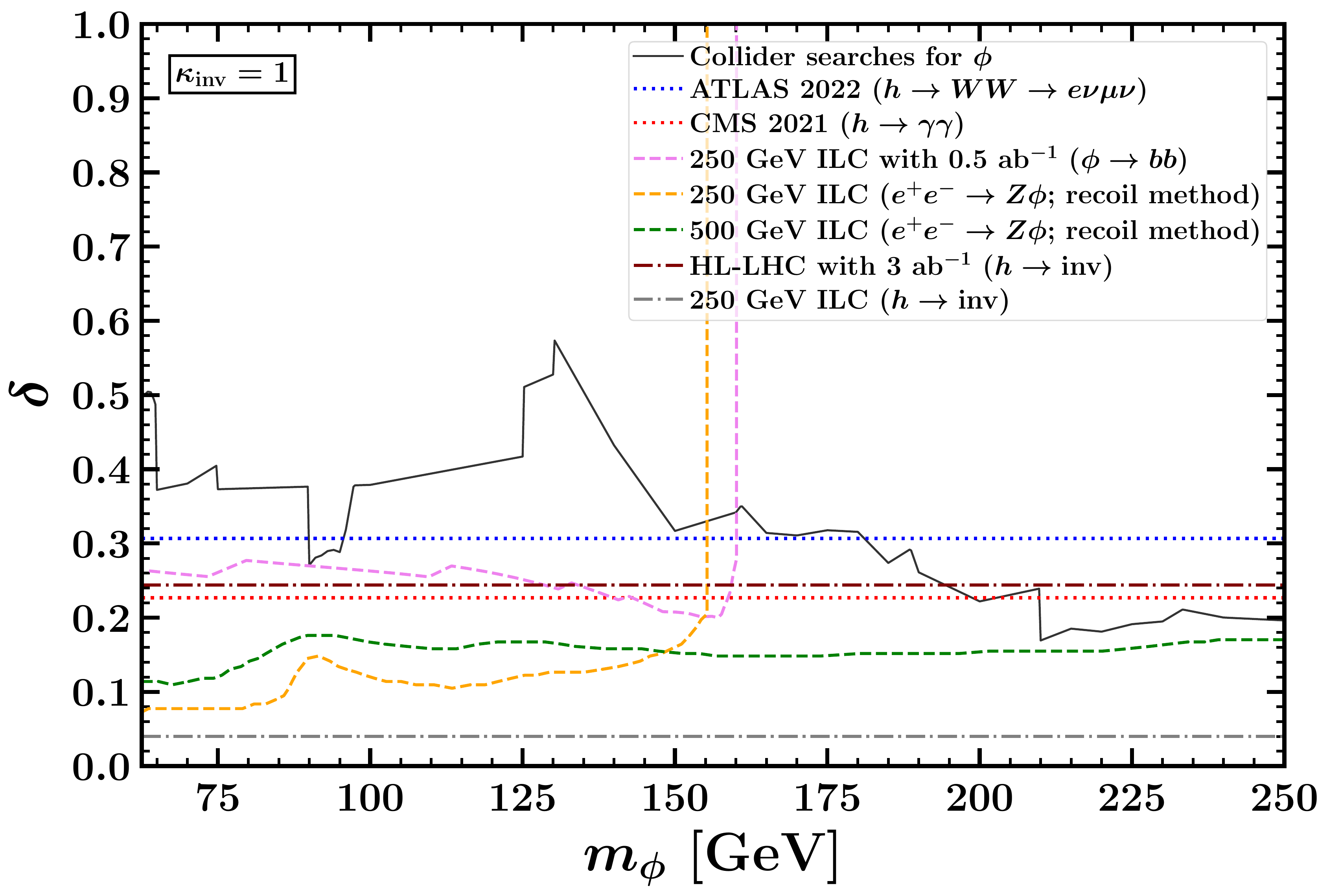}
  \end{minipage}
 \caption{\label{fig:phi_ILCproj}Some future sensitivities for $\delta$
as a function of the real singlet scalar mass $m_\phi$ for $\kappa_{\rm inv} = 0$
(left panel) and $\kappa_{\rm inv} = 1$ (right panel).
The solid and dotted lines show the present collider constraints from the searches for
$\phi$ and precision probes for $h$ respectively (see Figures~\ref{fig:h_bounds} and \ref{fig:phi_bounds}).
The violet dashed lines show the projected sensitivity of the 250 GeV ILC for $\phi \rightarrow b \overline{b}$ searches with some modest running assumptions \cite{Drechsel:2018mgd}
as discussed in the text.
The (orange, green) dashed lines show the model-independent expected sensitivity at the ILC with $\sqrt{s}=(250, 500)$ GeV, respectively, with various running assumptions \cite{Wang:2020lkq} as mentioned in the text.
The (maroon, gray) dash-dotted lines in the right panel are future projections for invisible searches of the observed Higgs boson $h$ from Figure~\ref{fig:h_bounds} at (HL-LHC with 3 ab$^{-1}$ \cite{Liu:2016zki}, 250 GeV ILC with 1.8 ab$^{-1}$ \cite{Potter:2022shg}), respectively, that indirectly constrain the parameter space.}
\end{figure}
%%%%%%%%%%%%%%%%%%%%%%%%%%%%%%%%%%%%%%%%%%%%%%%%%%%%%%%%%%%%%%%%%%%%%%%%%%%%%%%%%%

In Figure~\ref{fig:phi_ILCproj}, we show some future projections for $\delta$
as a function of the real singlet scalar mass $m_\phi$ for $\kappa_{\rm inv} = 0, 1$.
Here we only show the sensitivities of ILC for $\phi$ from
the $\phi \rightarrow b \overline{b}$ decays and
using a model-independent recoil mass method in $e^+ e^- \rightarrow Z \phi$
process \cite{Drechsel:2018mgd,Wang:2020lkq,Robens:2022erq}.
The solid and dotted lines show the present collider constraints from the searches for
$\phi$ and precision probes for $h$ respectively (see Figures~\ref{fig:h_bounds} and \ref{fig:phi_bounds}).
The violet dashed lines show the projected sensitivity of the ILC with $\sqrt s = 250$ GeV for $\phi \rightarrow b \overline{b}$ searches \cite{Drechsel:2018mgd} with 0.5 ab$^{-1}$ and 
beam polarization of $(80, 30)$\% for $(e^-, e^+)$ respectively.
The orange dashed lines show the expected sensitivity,
obtained using the model-independent recoil mass method in $e^+ e^- \rightarrow Z \phi$
process,
of the 250 GeV ILC with integrated luminosity of $(0.9, 0.9, 0.1, 0.1)$ ab$^{-1}$
for $(e^-_L e^+_R, e^-_R e^+_L, e^-_L e^+_L, e^-_R e^+_R)$ respectively and 
beam polarization of $(80, 30)$\% for $(e^-, e^+)$ respectively \cite{Wang:2020lkq}.
And, the green dashed lines show similar results but with 500 GeV ILC with integrated luminosity of
$(1.6, 1.6, 0.4, 0.4)$ ab$^{-1}$
for $(e^-_L e^+_R, e^-_R e^+_L, e^-_L e^+_L, e^-_R e^+_R)$ respectively \cite{Wang:2020lkq}.
Both orange and green dashed lines are independent of the choice for $\kappa_{\rm inv}$
as the mass of $\phi$ can be measured using the recoil mass against the $Z$ boson (reconstructed from $\mu^+ \mu^-$) independent of the decays of $\phi$.
The dash-dotted lines in the right panel with $\kappa_{\rm inv} = 1$ are the future projections for invisible searches of the observed Higgs boson
$h$ at HL-LHC and 250 GeV ILC (see Figure~\ref{fig:h_bounds}) that indirectly constrain the parameter space. 

To summarize, we recasted the bounds/projections reported
in various collider searches/studies, for neutral Higgs boson(s),
for the depletion factors $(\delta, \kappa_{\rm inv})$
of the already discovered 125 GeV Higgs boson $h$ that occur in the SM extended
with a real scalar singlet. And, we found that the precision studies of the observed Higgs boson $h$
often provide for the strongest constraint/reach in the parameter space
$(\delta, \kappa_{\rm inv}, m_{\phi})$ compared to the direct searches for the additional
physical Higgs $\phi$, at least for $m_h/2 \le m_\phi \le 2 m_h$.
For large $m_\phi$, we found that the precision electroweak observables $S, T,$ and $U$
can impose a strong constraint on the parameter space.
The results obtained in this section for the real singlet scalar extension suggest that in
various other extended Higgs sectors that contain ``the depleted Higgs boson",
the precision probes for the 125 GeV Higgs boson alone can often best constrain such a
class of models.

%%%%%%%%%%%%%%%%%%%%%%%%%%%%%%%%%%%%%%%%%%%%%%%%%%%%%%%%%%%%%%%%%%%%%%%%%%%%%%%%%%
%%%%%%%%%%%%%%%%%%%%%%%%%%%%%%%%%%%%%%%
\section{High multiplicity of real singlet scalars\label{sec:NSinglets_Higgs_bounds}}
%\subsection{$N$ real singlet scalars \label{sec:NSinglets_Higgs_bounds}}
\setcounter{equation}{0}
\setcounter{figure}{0}
\setcounter{table}{0}
\setcounter{footnote}{1}

Finally, we also briefly consider an extension to the SM with $N$ real SM gauge-singlet scalars $S_{i = 1, 2, \ldots, N}$, each with an invisible width $\Gamma_{\rm inv}$, where the gauge basis $\{h_{\rm  SM}, S_i\}$ is related to the mass basis $\{h, \phi_i\}$ in the following way:
\beq
\begin{pmatrix}
h_{\rm SM}\\
S_1\\
S_2\\
.\\
.
\end{pmatrix}
&=&
\begin{pmatrix}
\sqrt{1 - \delta^2} & \eta & \eta & ~.~ & ~.~ \\
-\eta & 1 + \epsilon & \epsilon & ~.~ & ~.~ \\
-\eta & \epsilon & 1 + \epsilon & ~.~ & ~.~ \\
. & . & . & . & .\\
. & . & . & . & .
\end{pmatrix}
\begin{pmatrix}
h\\
\phi_1\\
\phi_2\\
.\\
.
\end{pmatrix}
.
\eeq
Here, for purposes of illustration, we chose that the gauge-singlets $S_i$ mix equally among themselves (parameterized by $\epsilon$) and
with the SM Higgs $h_{\rm SM}$ (parameterized by $\eta$).
Note that $h$ here stands for the physical Higgs boson with mass $\sim 125$ GeV that is already discovered, and $\phi_i$ are the exotic Higgs with masses taken to be $m_h/2 \leq m_{\phi_i} \leq 2 m_h$ for simplicity.
The orthogonality of the above matrix implies that
\beq
\eta &=& \frac{\delta}{\sqrt N},\\
\epsilon &=& \frac{1}{N} \left ( \sqrt{1 - \delta^2} - 1 \right ) = - \frac{\delta^2}{2 N} + \ldots .
\eeq
The gauge eigenstates can be expressed in terms of the mass eigenstates:
\beq
h_{\rm SM} &=& \sqrt{1 - \delta^2} \, h + \frac{\delta}{\sqrt N} \sum_{i=1}^N \phi_i,\\
S_i &=& -\frac{\delta}{\sqrt N} \, h + \phi_i + \frac{1}{N} \left ( \sqrt{1 - \delta^2} - 1 \right ) \sum_{j=1}^N \phi_j
\eeq
%%%%%%%%%%%%%%%%%%%%%%%%%%%%%%%%%%%%%%%%%%%%%%%%%%%%%%%%%%%%%%%%%%%%%%%%%%%%%%%%%%
\begin{figure}
  \begin{minipage}[]{0.495\linewidth}
    \includegraphics[width=8cm]{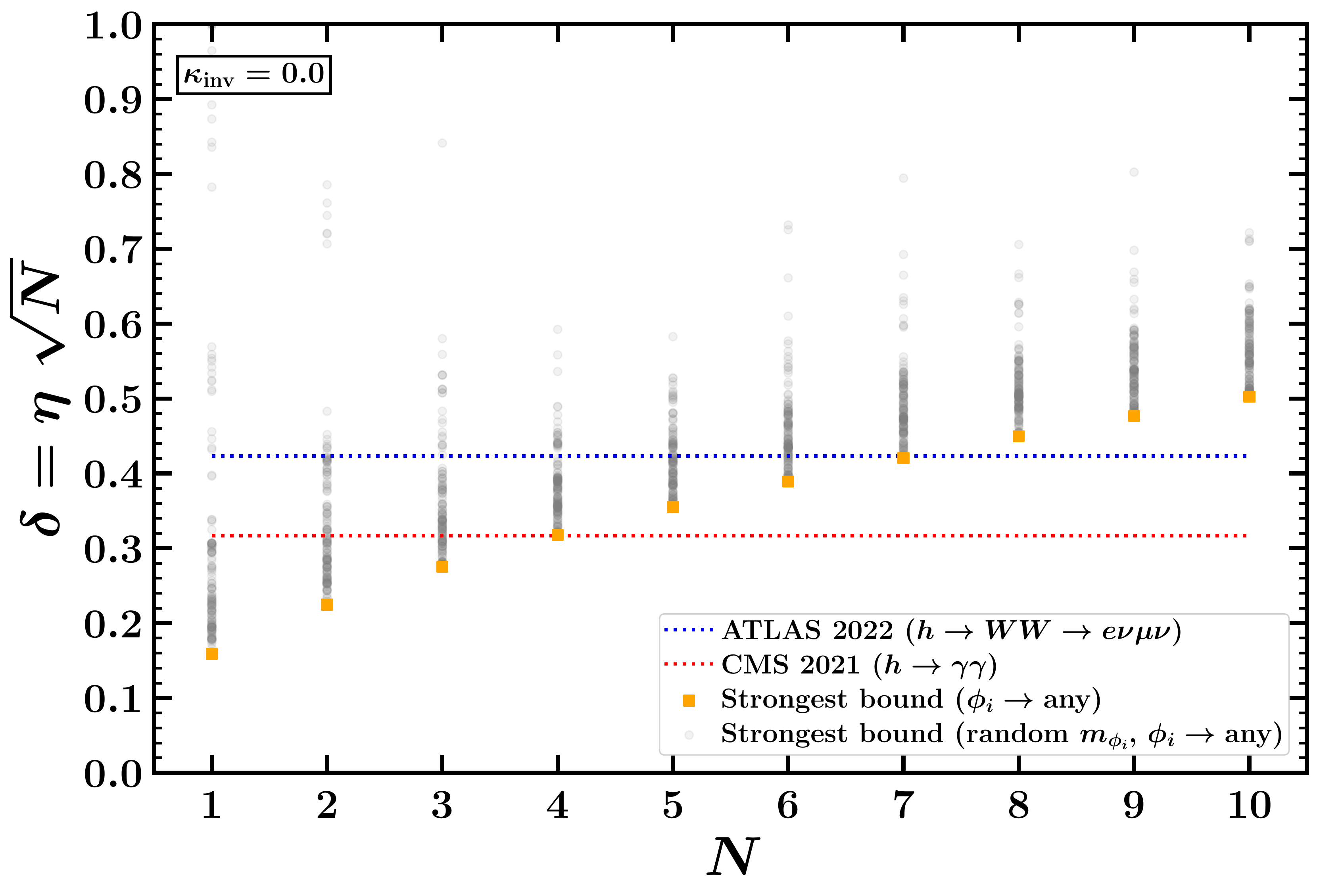}
  \end{minipage}
  \begin{minipage}[]{0.495\linewidth}
    \includegraphics[width=8cm]{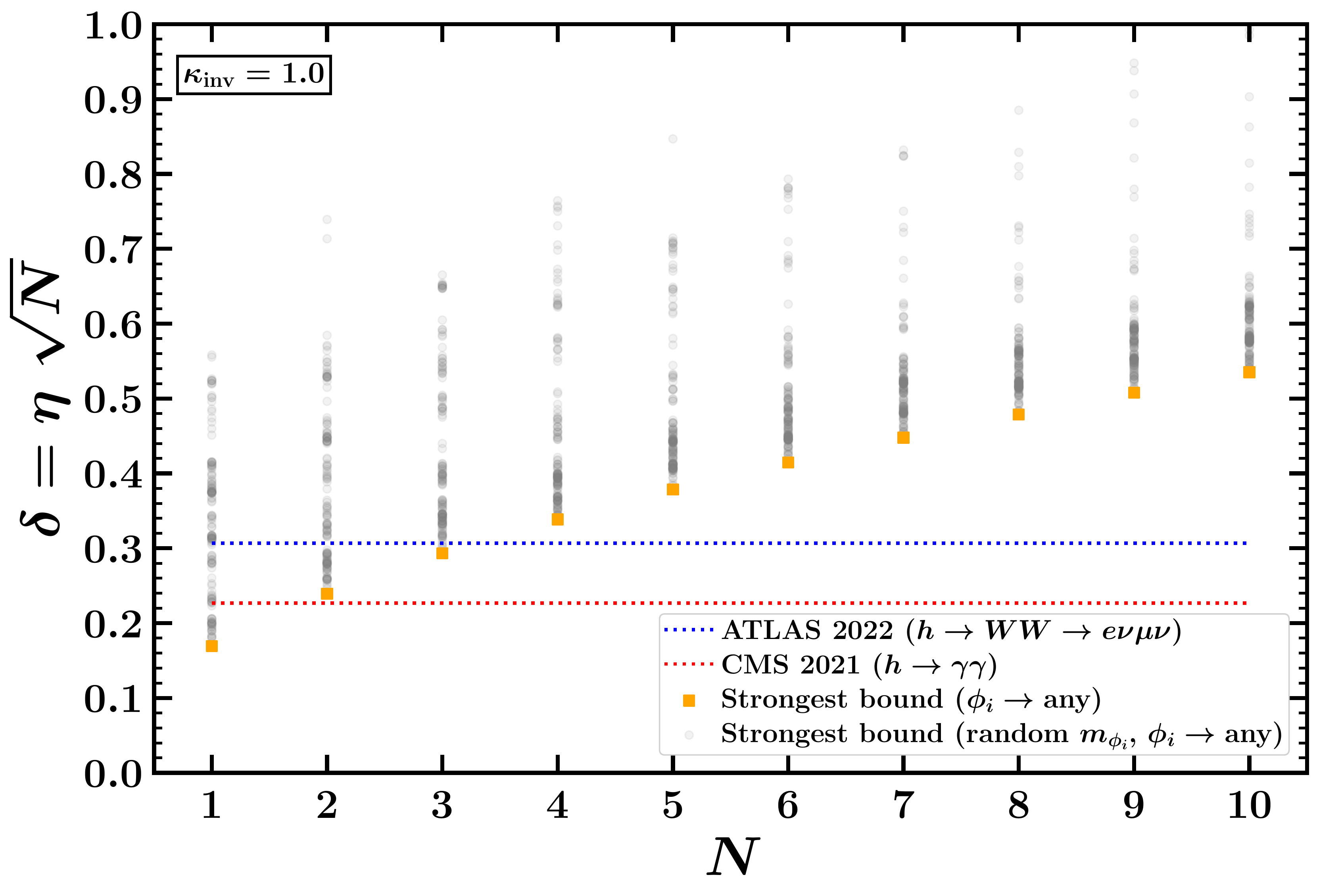}
  \end{minipage}
 \caption{\label{fig:deltavsN}Constraints on $\delta$ for the SM extended with $N$ real singlet scalars for $\kappa_{\rm inv} = 0$ (left panel) and $\kappa_{\rm inv} = 1$ (right panel).
 Dotted lines show the strongest bounds from the precision probes for the 125 GeV Higgs $h$,
 scatter plot with square markers show the strongest bound from the searches for
 the exotic Higgs bosons $\phi_i$ obtained by varying the mass of one of the exotic Higgs $m_{\phi_i}$ from $m_h/2$ to $2 m_h$.
 And the light-gray scatter plots show the strongest constraints in various iterations where
 the masses $m_{\phi_{i = 1, \ldots N}}$ are randomly distributed between $m_h/2$ and $2 m_h$, illustrating that the precision probes for the 125 GeV Higgs boson $h$ alone typically provide for the strongest constraints.}
\end{figure}
%%%%%%%%%%%%%%%%%%%%%%%%%%%%%%%%%%%%%%%%%%%%%%%%%%%%%%%%%%%%%%%%%%%%%%%%%%%%%%%%%%
%%%%%%%%%%%%%%%%%%%%%%%%%%%%%%%%%%%%%%%%%%%%%%%%%%%%%%%%%%%%%%%%%%%%%%%%%%%%%%%%%%
\begin{figure}
  \begin{minipage}[]{0.95\linewidth}
    \includegraphics[width=15cm]{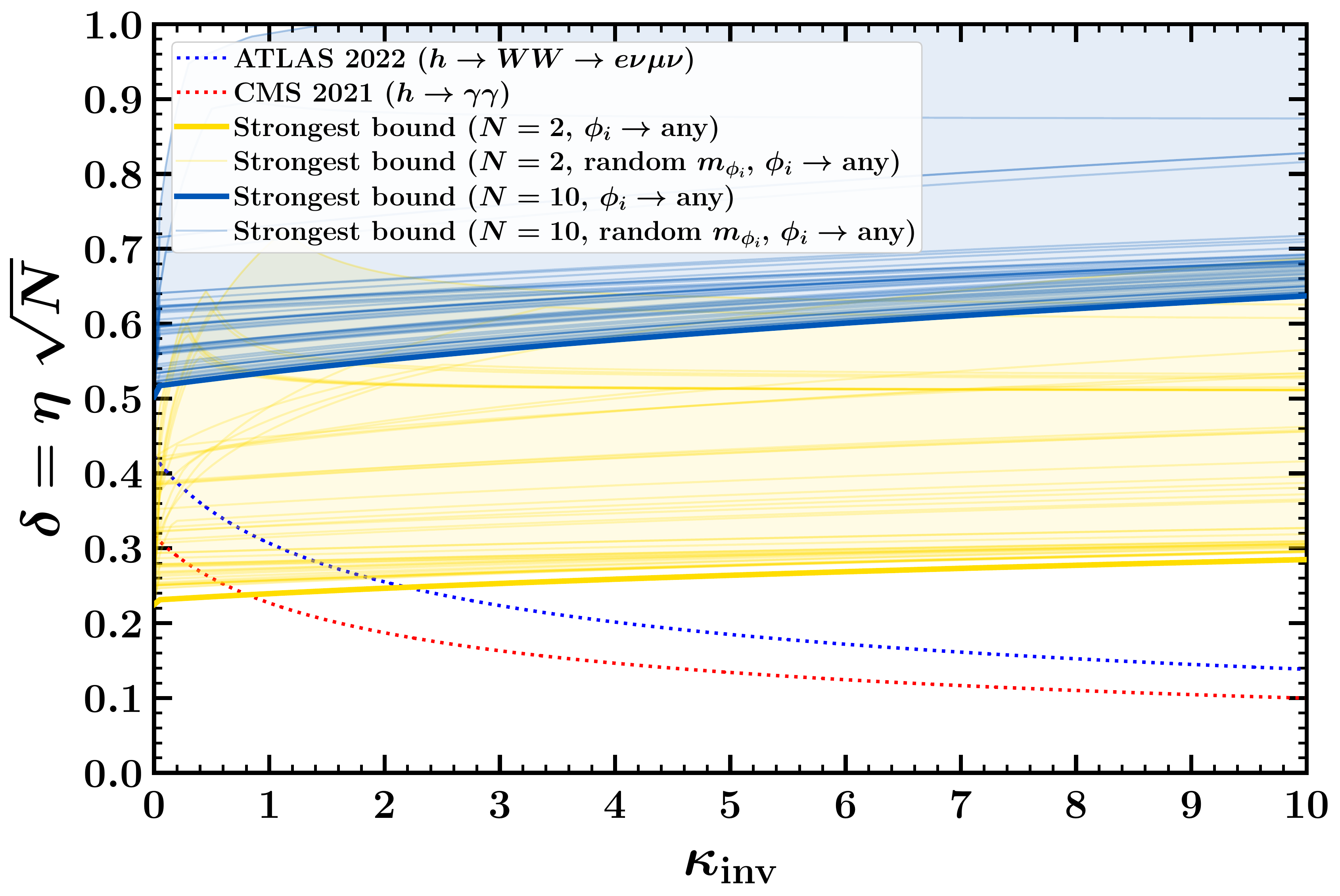}
  \end{minipage}
 \caption{\label{fig:deltavskappa_N}Constraints on $\delta$ for the SM extended with $N$ real singlet scalars as a function of $\kappa_{\rm inv}$.
 The dotted lines show the strongest bounds from the (visible) precision probes for the 125 GeV Higgs $h$,
 and the thicker solid lines show the strongest bound from the searches for
 the exotic Higgs bosons $\phi_i$ obtained by varying the mass of one of the exotic Higgs $m_{\phi_i}$ from $m_h/2$ to $2 m_h$ for $N=2$ (blue) and $N=10$ (yellow).
 And the thinner solid lines show the strongest constraints in various iterations where
 the masses $m_{\phi_{i = 1, \ldots N}}$ are randomly distributed between $m_h/2$ and $2 m_h$, illustrating that the precision probes for the 125 GeV Higgs boson $h$ alone typically provide for the strongest constraints.}
\end{figure}
%%%%%%%%%%%%%%%%%%%%%%%%%%%%%%%%%%%%%%%%%%%%%%%%%%%%%%%%%%%%%%%%%%%%%%%%%%%%%%%%%%

We once again find that the total widths, branching ratios, and production cross-sections of $h$ are same as in
Section~\ref{sec:125GeVHiggs}, and for the exotic Higgs bosons $\phi_i$, the results for $\phi$ from Section~\ref{sec:scalar_ext} are directly applicable but with the replacement $\delta \rightarrow \eta = \frac{\delta}{\sqrt N}$, for example:
\beq
\frac{\Gamma^{\phi_i}}{\Gamma_{\rm SM}} &=&
\frac{\delta^2}{N} + \left ( 1 - \frac{\delta^2}{N} \right ) \frac{\kappa_{\rm inv}}{\kappa_{\rm SM}}.
\eeq
Therefore all the bounds in Section~\ref{sec:Singlet_Higgs_bounds} can be reinterpreted for each $\phi_i$ by simply rescaling the bounds by $\sqrt N$.
We can then immediately see that the bounds coming from the searches for $\phi_i$ become weaker as $N$ gets larger,
and the strongest constraints often come directly from the precision probes for the 125 GeV Higgs boson $h$.
To illustrate this, we randomly generate $m_{\phi_i} \in [m_h/2, 2 m_h]$ and choose the strongest bound from the bounds obtained for each $m_{\phi_i}$. We then iterate these steps several times, so that it is evident that the strongest bound always (often) comes from the precision probes for the 125 GeV Higgs boson $h$ for large (small) $N$.

Figure~\ref{fig:deltavsN} shows various bounds on $\delta$ as a function of number of SM gauge-singlet
real scalars $N$ for fixed $\kappa_{\rm inv}$. Dotted lines show the constraints coming from the precision probes for the 125 GeV Higgs boson $h$, specifically the most stringent ones from the precision probes for $h \rightarrow \gamma \gamma$ by CMS and $h \rightarrow W W \rightarrow e \nu \mu \nu$ by ATLAS.
The orange-colored square scatter plot shows the strongest bound on
$\delta$ from searches for $\phi_i$ for all $m_{\phi_i} \in [m_h/2, m_h]$.
For example, for $\kappa_{\rm inv} = 0$, the strongest bound on $\eta$ can be directly read-off of the top-left plot of Figure~\ref{fig:phi_bounds} to be $\sim 0.16$ at $m_{\phi_i} \sim 65$ GeV
and so the strongest bound on $\delta = 0.16 \sqrt N$.
Various light-gray scatter plots are the strongest bounds obtained by randomly distributing
all $m_{\phi_i}$ between 62.5 and 250 GeV.
And, in Figure~\ref{fig:deltavskappa_N}, we show various bounds on $\delta$ as a function of $\kappa_{\rm inv}$.
Dotted lines, once again, show the bounds from the precision probes for the observed Higgs boson $h$.
Thick solid lines show the strongest bounds from the searches for $\phi_i$
(for all $m_{\phi_i}$) for real scalar singlet extensions with $N=2$ and $N=10$ scalars.
Each thin solid line shows the strongest bounds obtained by randomly choosing $m_{\phi_i}$ in between $m_h/2$
and $2 m_h$.
As mentioned before, it is clear from Figures~\ref{fig:deltavsN} and \ref{fig:deltavskappa_N} that
the strongest bounds almost always come from the precision probes for the 125 GeV Higgs boson $h$ for large $N$ and/or large $\kappa_{\rm inv}$.

Therefore, as remarked at the end of the previous section, we found
that the indirect precision probes for the observed Higgs boson with $m_h \sim 125$ GeV tend to
give the most powerful constraint on the depletion factors $(\delta, \kappa_{\rm inv})$
in the SM extended with a high multiplicity of real singlet scalars,
albeit under some simplifying assumptions. And, the constraints from the direct searches
for the additional exotic Higgs states $\phi_i$ get weaker as
the multiplicity or the invisible width of the scalars increase.

%%%%%%%%%%%%%%%%%%%%%%%%%%%%%%%%
%%%%%%%%%%%%%%%%%%%%%%%%%%%%%%%%
%%%%%%%%%%%%%%%%%%%%%%%%%%%%%%%%
\section{Conclusion \label{sec:conclusion}}
\setcounter{equation}{0}
\setcounter{figure}{0}
\setcounter{table}{0}
\setcounter{footnote}{1}

In this paper, we obtained the latest bounds on the universal depletion factor $\delta$
that suppresses all the couplings of the
observed Higgs boson $h$ to SM final states from the recent LHC searches in various decay channels as
a function of  the other depletion factor $\kappa_{\rm inv}$, a parameter related to the invisible width of $h$ (see eq.~(\ref{eq:kappainv})).
We argued that these bounds indirectly constrain many extended Higgs sectors that give rise to $\delta$
and $\kappa_{\rm inv}$, and are also in general comparable to or stronger than the direct searches for the
additional Higgs states, at least if their masses are in between $m_h/2$ and $2 m_h$.

To demonstrate we considered various bounds that come from the collider searches for a exotic Higgs boson $\phi$ that occurs in the SM extended with a real SM gauge-singlet scalar
that decays only invisibly to some exotic hidden sector fermions. We also obtained the constraints
on $\delta$ as a function of the mass of the exotic Higgs from precision electroweak observables
(in particular the Peskin-Takeuchi parameters). Although, the constraints from the $S, T,$ and $U$
parameters are not the strongest bounds for $m_h/2 \le m_\phi \le 2 m_h$, these bounds get very strong
for higher $m_\phi$.
And, moreover, we also considered some future sensitivities for the observed Higgs boson $h$ and
for the exotic Higgs boson $\phi$, to show that the parameter space of the extended Higgs sectors,
that lead to the universal depletion factor $\delta$ and an invisible width factor $\kappa_{\rm inv}$
for $h$, get more constrained and will eventually measure the deviations from the SM predictions and
lead the way to the discovery of the additional Higgs states.

Finally, we also considered the case where there are actually $N$ real singlet scalars,
with some simplifying assumptions for illustration purposes,
and found that the indirect bounds from the precision probes for the $125$ GeV Higgs boson $h$ alone tend
to moderately constrain such scenarios. The cases with larger number of such scalars and/or
larger invisible widths for $h$ are much more constrained from the indirect searches.

{\it Acknowledgments:}
We thank John Thiels for his help and support in using
the Great Lakes cluster at University of Michigan.
This research is supported in part through computational resources and services provided
by Advanced Research Computing (ARC), a division of Information and Technology Services
(ITS) at the University of Michigan, Ann Arbor.
This work is supported in part by the Department of Energy under grant number DE-SC0007859.
%%%%%%%%%%%%%%%%%%%%%%%%%%%%%%%%%%%%%%%%%%%%%%%%%%%%%%%%%%%%%%%%%%%%%%%%

\end{document}